\documentclass[aps,prc,twocolumn,floatfix,superscriptaddress]{revtex4}

\usepackage{epsfig}
\usepackage{amsmath}

\usepackage{graphicx}

\begin{document}

\title{Single photons from relativistic collision of lead nuclei at CERN SPS: A
reanalysis}  
\author{Rupa Chatterjee}
\affiliation{Variable Energy Cyclotron 
Centre, 1/AF Bidhan Nagar, Kolkata 700 064, India}            
\author{Dinesh K.~Srivastava}
\affiliation{Variable Energy Cyclotron 
Centre, 1/AF Bidhan Nagar, Kolkata 700 064, India}            
\author{Sangyong Jeon}
\affiliation{Department of Physics, McGill University, Montr\'eal,
 Canada H3A 2T8}

\begin{abstract}
We present a reanalysis of single photon production from 
relativistic collision of lead nuclei at CERN SPS measured 
by the WA98 experiment. The refinements include use of 
iso-spin and shadowing corrected NLO pQCD treatment for 
prompt photon production using an optimized scale for 
factorization, renormalization, and fragmentation and use 
of hydrodynamics suited for non-central collisions along 
with a well tested equation of state admitting a quark-hadron 
phase transition. A quantitative explanation of the data
requires a large initial temperature (at a small formation
time of about 0.2 fm/$c$)  and a moderate increase in the 
prompt yield which could perhaps be attributed to the Cronin 
effect in nuclei. The data can also be explained using a moderate 
initial temperature (at a formation time of about 1 fm/$c$) with 
a very large $K$-factor multiplying the prompt yield.  We show 
that different initial times give rise to different values for 
the elliptic flow parameter $v_2$ for thermal photons. We also 
show that a measurement of $v_2$ for thermal photons could also 
distinguish between the scenarios with or without a phase 
transition. 
\end{abstract}

\pacs{25.75.-q,12.38.Mh}
\maketitle
\section{Introduction}
The first observation of single photons in relativistic 
collision of lead nuclei in the WA98 experiment at CERN 
SPS~\cite{wa98} remains an important mile-stone in our 
search for  the quark-hadron phase transition. The earlier 
experiment studying the $S+Au$ collisions had provided only 
(though quite useful) upper limit on the single photon 
production~\cite{wa80}. The importance of the single photons 
stems from the expectation that once produced they leave the 
system without any further interaction (see Ref~\cite{dks_qm08}, 
for a recent account of nodal developments in this field). It 
is thus, expected that if a thermalized system of quarks and 
gluons or hot hadrons is produced in such collisions, its 
temperature could be related to the spectrum of the single 
photons. On the experimental front, the success of this 
endeavour hinges on our ability to subtract out the decay 
photons from the inclusive spectrum of photons, while on the 
theoretical front it depends on our ability to evaluate 
non-thermal photons in a quantitative manner.

In the present work, we re-analyze the single photon 
measurements reported by the WA98 experiment after 
incorporating several recent improvements in our
understanding of the physics of relativistic heavy 
ion collisions. Firstly, we perform the NLO pQCD 
evaluation of prompt photon production using the 
optimized scale for factorization, renormalization, 
and fragmentation, $Q=p_T/2$ which has been found 
to describe a vast body of single photon production 
in $pp$ collisions without introduction of any
%
%SY: added Aurenche's refs
%
intrinsic $k_T$ for protons~\cite{pat1,pat2}.
We explicitly account for the iso-spin of the projectile 
target system, which affects the results at large $x_T=2 
p_T/\sqrt{s}$ and include the effects of parton shadowing~\cite{eks98}. 
Next we account for the azimuthal anisotropy of the system 
for non-central collisions, while performing the hydrodynamics 
calculations. Finally we explore the effects of varying the 
initial conditions to set limits on the likely initial temperature.

In the next section we briefly discuss our estimates for the 
prompt photon production. In Sect. III we describe the setting 
up of the initial conditions, and in Sect. IV we discuss the 
results. Finally we give our conclusions.

\begin{figure}
\centerline{\includegraphics*[width=5.5cm,angle=-90]{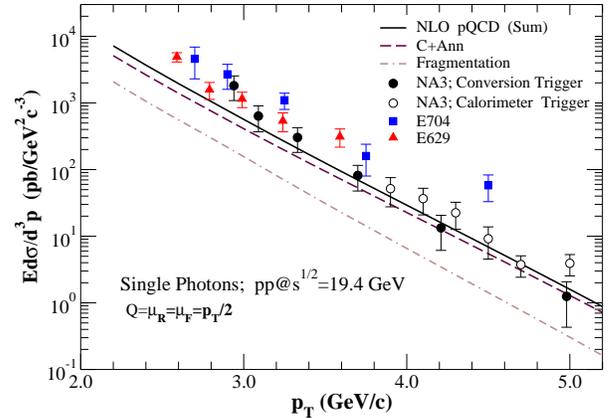}}
\caption{Prompt photons from $pp$ collisions at $\sqrt{s}=$ 19.4 
GeV. Experimental results for E704~\cite{e704} for $pp$ collisions 
and those estimated from $p+^{12}C$ collisions by the 
E629~\cite{e629} and NA3~\cite{na3} experiments are also 
given for a comparison. The results for the NA3 experiment 
use two different triggers; conversion and calorimeter. }
\label{fig1}
\end{figure}

\begin{figure}
\centerline{\epsfig{file=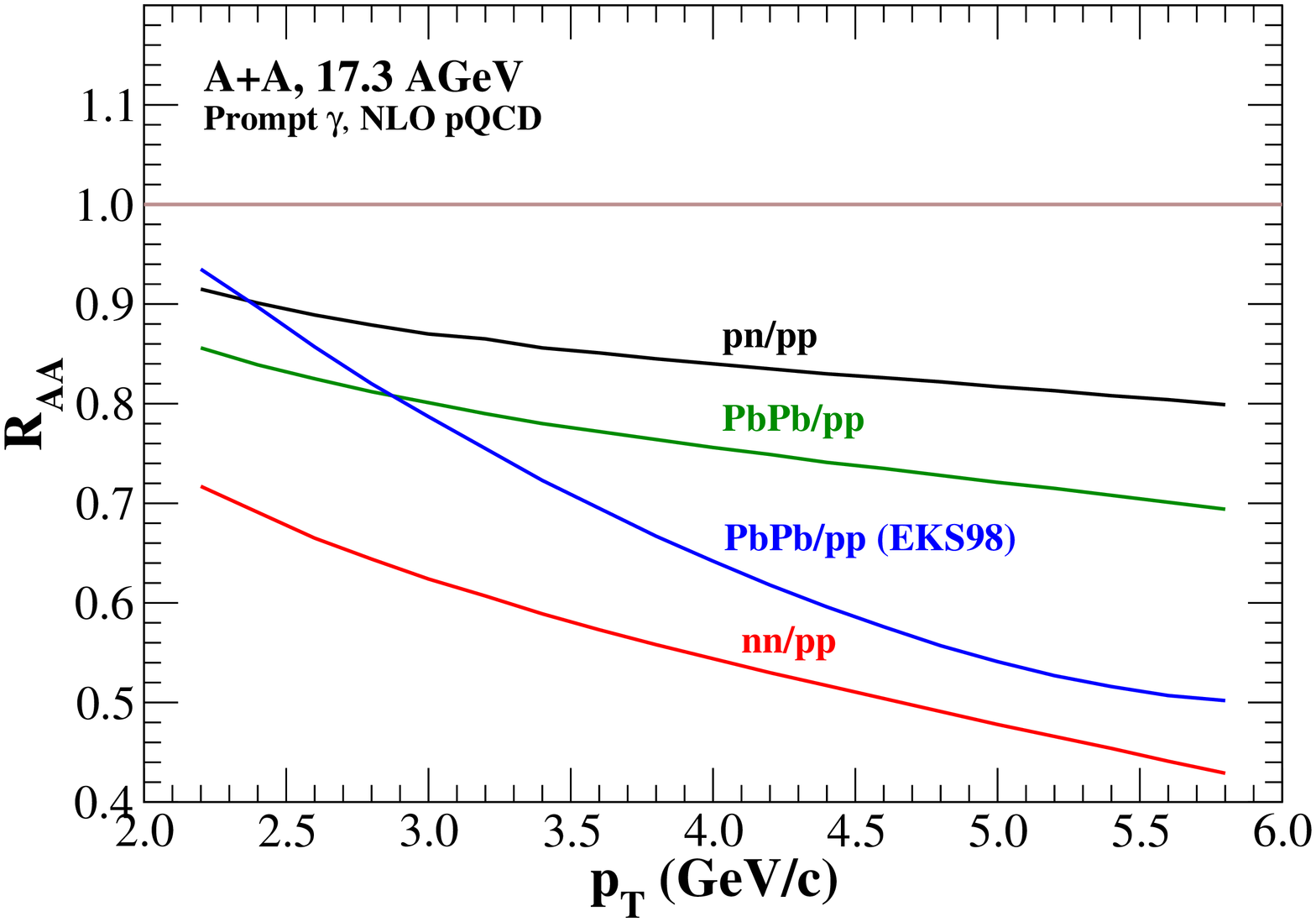,angle=-90,width=7.9cm}}

\centerline{\epsfig{file=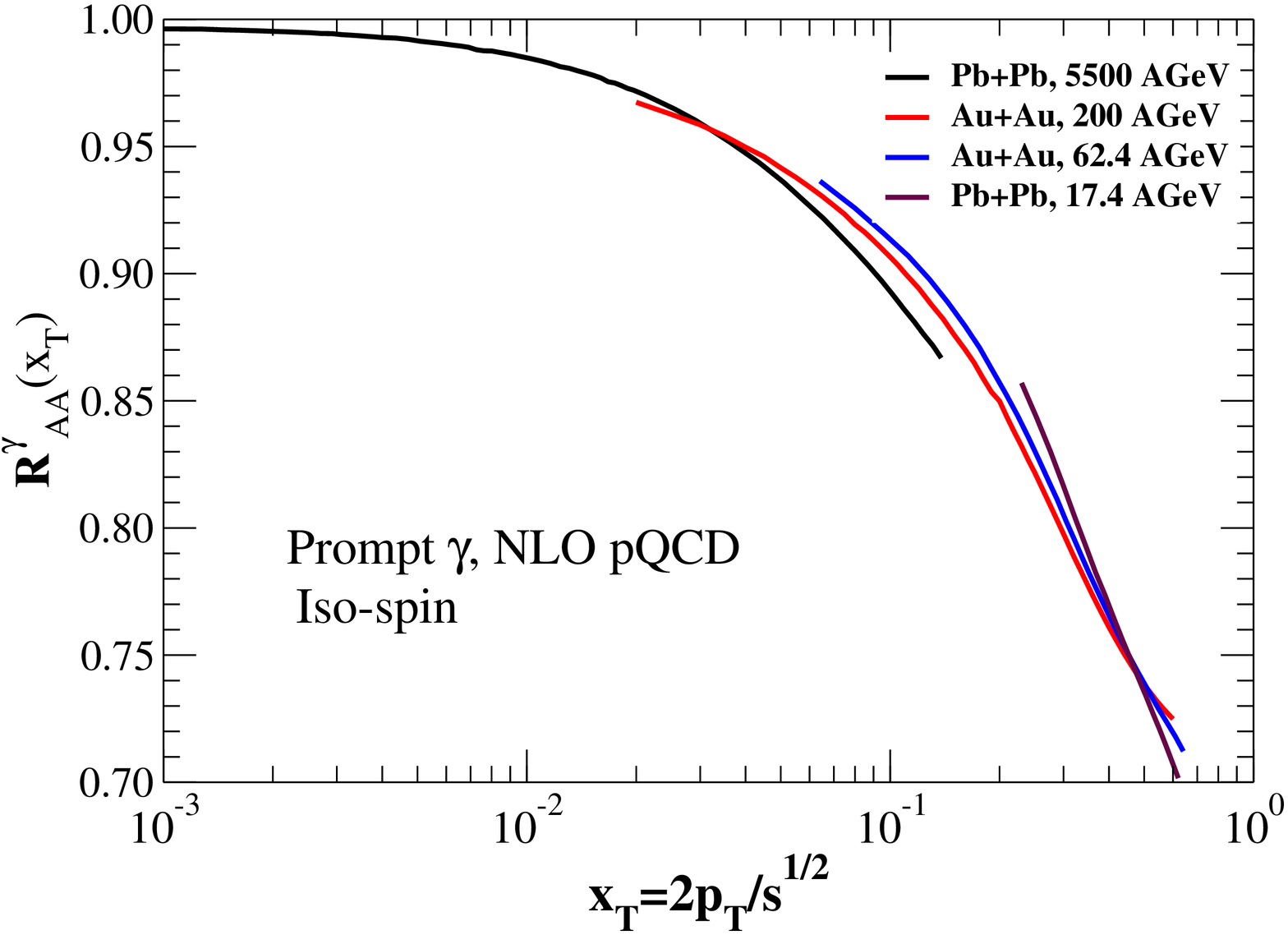,angle=-90,width=7.9cm}}

\centerline{\epsfig{file=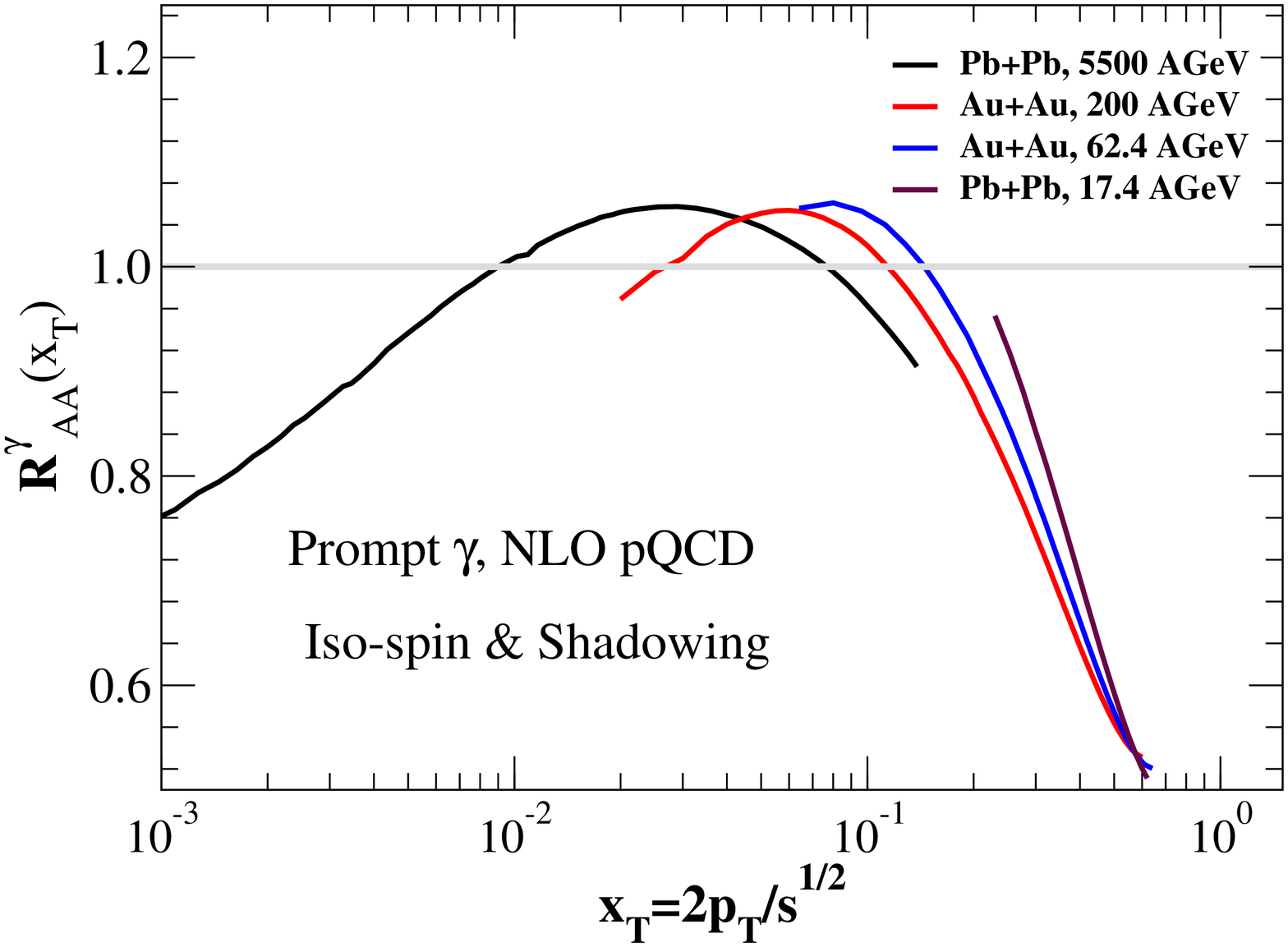,angle=-90,width=7.9cm}}

\caption{Upper panel: Effect of iso-spin and parton shadowing 
on production of prompt photons, calculated using NLO pQCD, at 
$\sqrt{s_{NN}}=$ 17.3 GeV, which corresponds to the nucleon-nucleon
centre of mass energy for the WA98~\cite{wa98} experiment.  Results 
are given in terms of the nuclear modification factor $R_{AA}$ for 
$pn$, $nn$ and $PbPb$ collisions. Middle panel: Effect of iso-spin 
at SPS, RHIC, and LHC energies as a function of $x_T=2p_T/\sqrt{s}$.
Lower panel: Same as before with shadowing. (see Ref.\cite{emprobes})}
\label{fig2}
\end{figure}

\begin{figure}[ht]
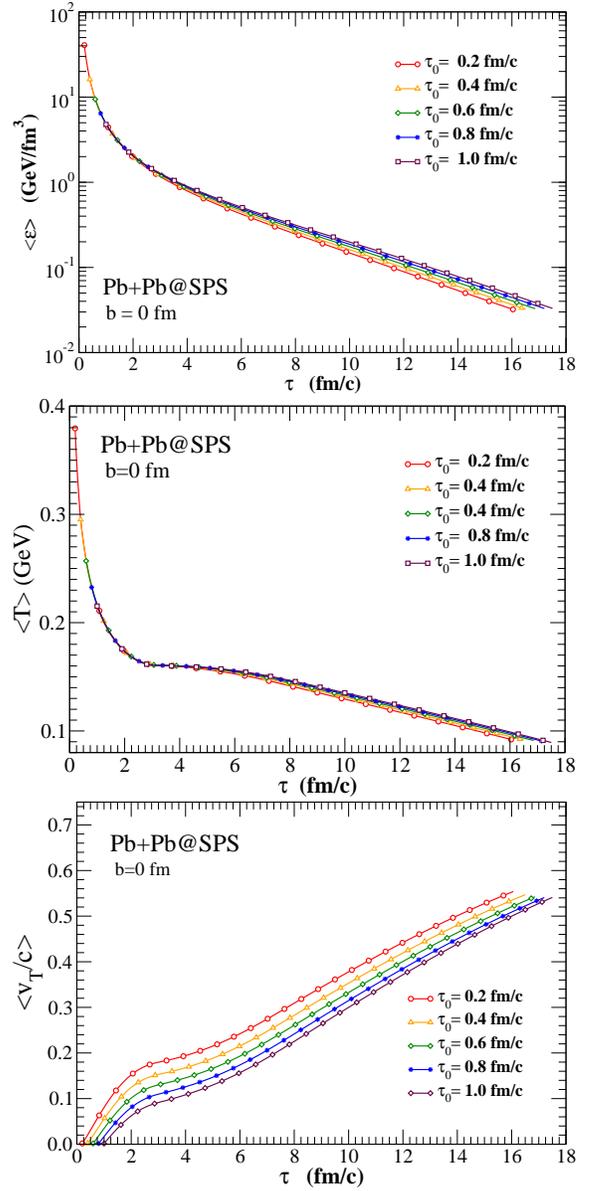

\centerline{\includegraphics[width=7.5cm,clip=ture]{eps.eps}}

\centerline{\includegraphics[width=7.5cm,clip=true]{temp.eps}}

\centerline{\includegraphics[width=7.5cm,clip=true]{vt.eps}}

\caption{Evolution of average energy density (upper panel), 
temperature (middle panel), and radial flow velocity (lower panel) 
with time for different initial times $\tau_0$ but identical 
rapidity density for total entropy and net baryons for a
central collision of two lead nuclei at SPS energies.}
\label{fig3}
\end{figure}

\begin{figure}[ht]
\centerline{\includegraphics[width=7.9cm]{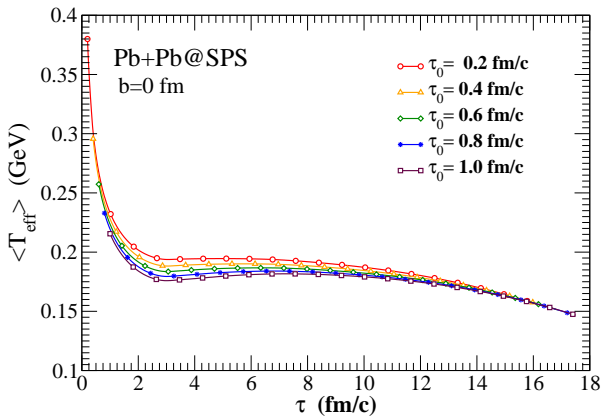}}
\caption{Evolution of average effective temperature
with time for different initial times $\tau_0$ but 
identical rapidity density for total entropy and net 
baryons for a central collision of two lead nuclei at 
SPS energies.}
\label{fig4}
\end{figure}

\begin{figure}[ht]
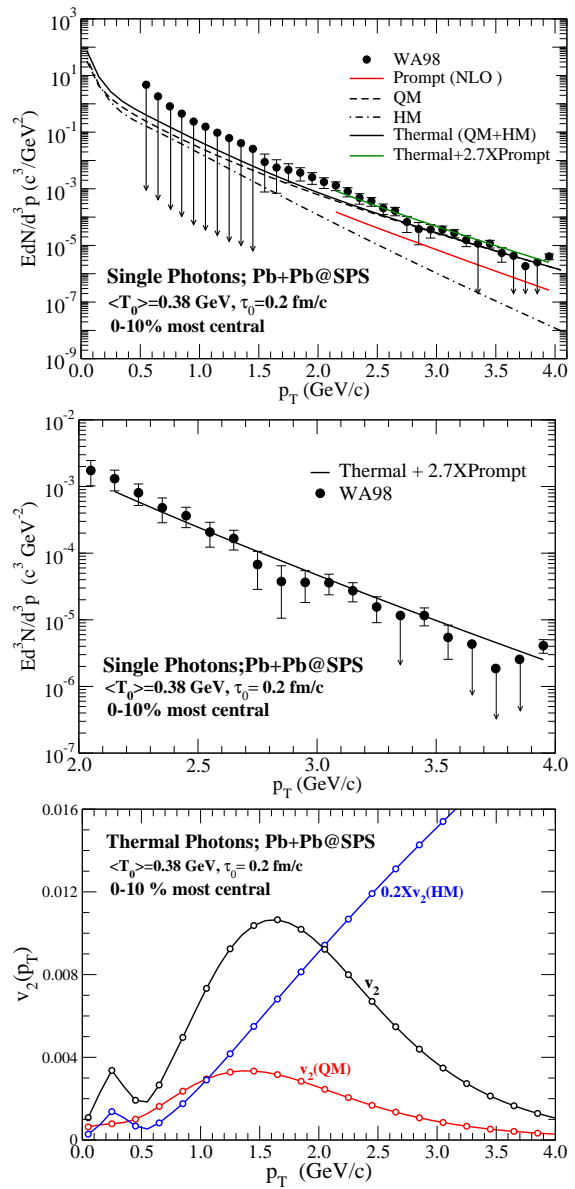

\centerline{\includegraphics*[width=7.4cm]{pb_0.2.eps}}

\centerline{\includegraphics*[width=7.4cm]{0.2.eps}}

\centerline{\includegraphics*[width=7.4cm]{0.2_v2.eps}}
\caption{Upper Panel: 
Fit to single photon spectra from Pb(158 AGeV)+Pb collisions 
measured by the WA98~\cite{wa98} experiment for $\tau_0=$ 
0.2 fm/$c$ The prompt photon contribution, is scaled by a 
factor of 2.7 to normalize the theoretical results to the 
experimental data at $p_T$ = 2.55 GeV/$c$. Middle panel: 
Details. Lower panel: Elliptical flow coefficients for the 
thermal photons. QM and HM stand for photons from quark matter 
and hadronic matter.}
\label{fig5}
\end{figure}

\begin{figure}[ht]
\centerline{\includegraphics*[width=7.5cm]{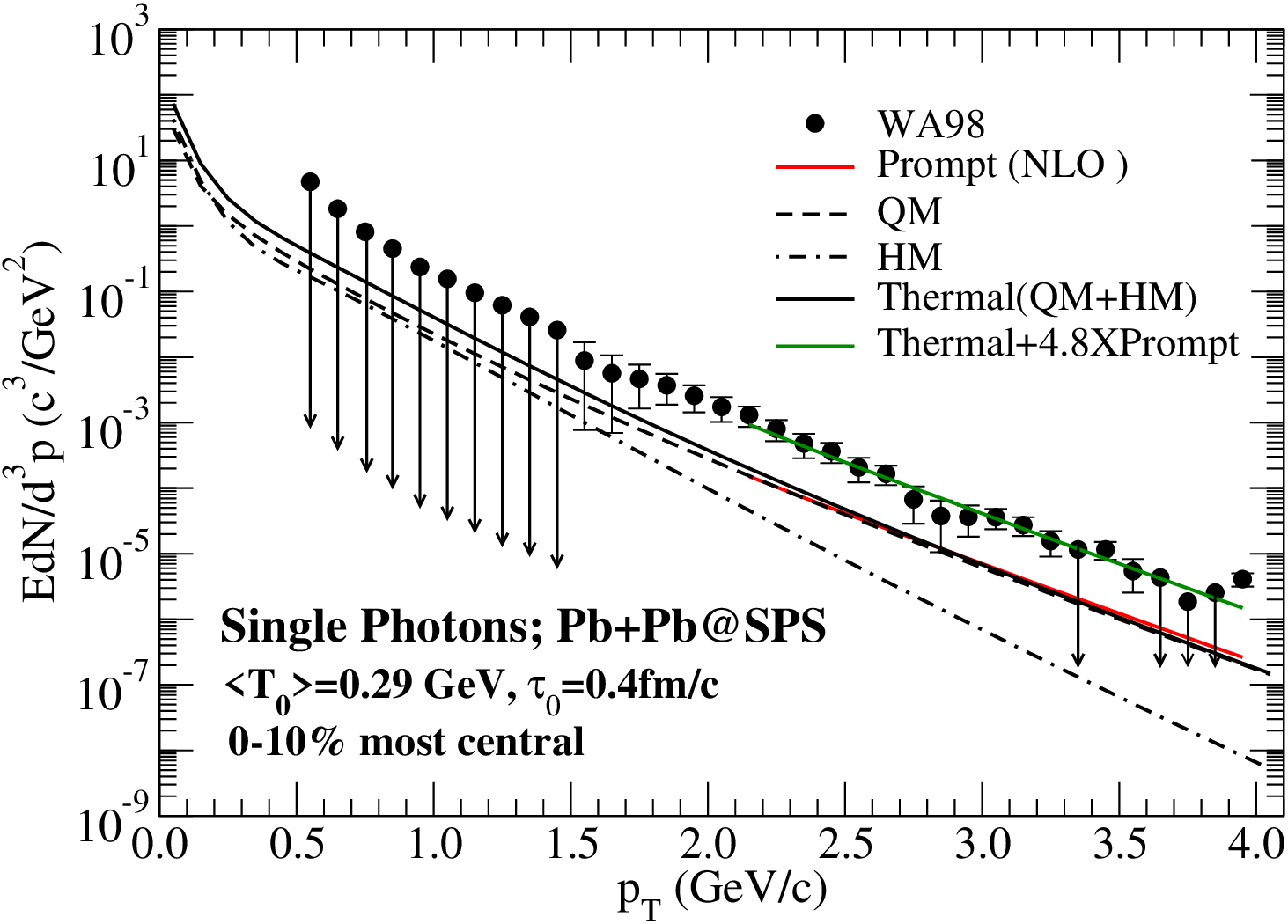}}
\centerline{\includegraphics*[width=7.5cm]{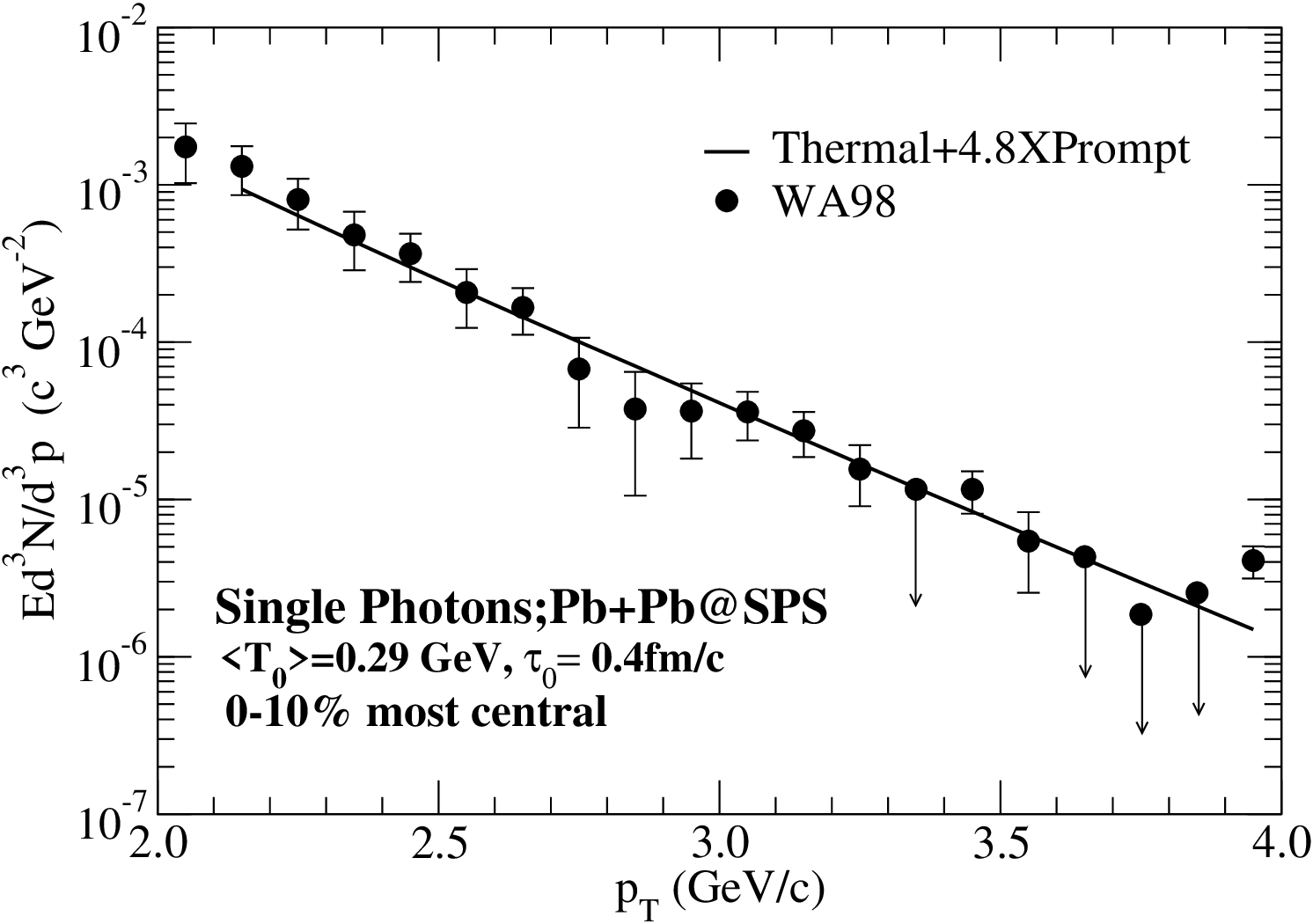}}
\centerline{\includegraphics*[width=7.5cm]{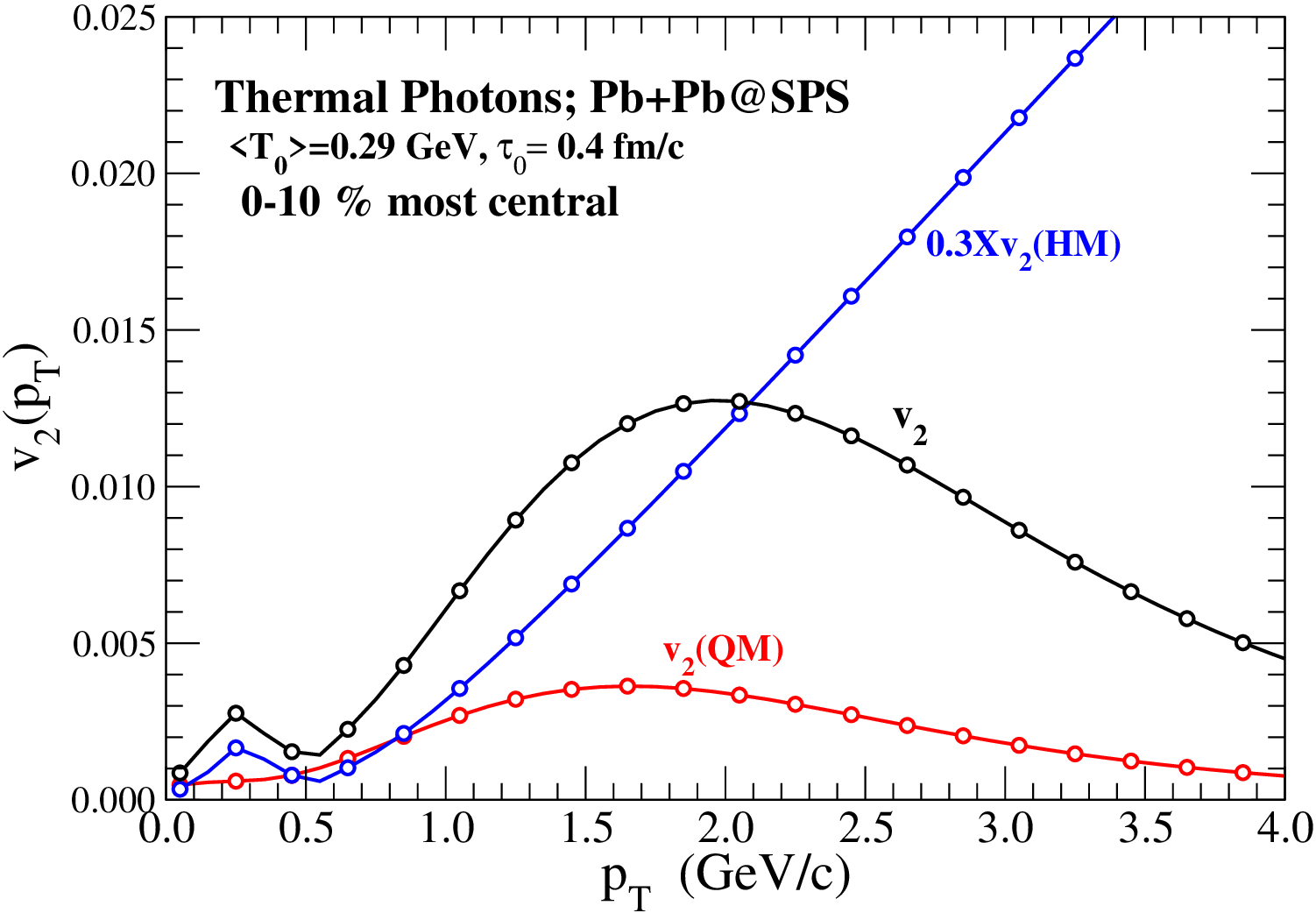}}
\caption{Same as Fig.~\ref{fig5} for $\tau_0$ = 0.4 fm/$c$. 
Note that the NLO pQCD results have to be scaled up by 4.8 
for describing the data} 
\label{fig6}
\end{figure}

\begin{figure}[ht]
\centerline{\includegraphics*[width=7.5cm]{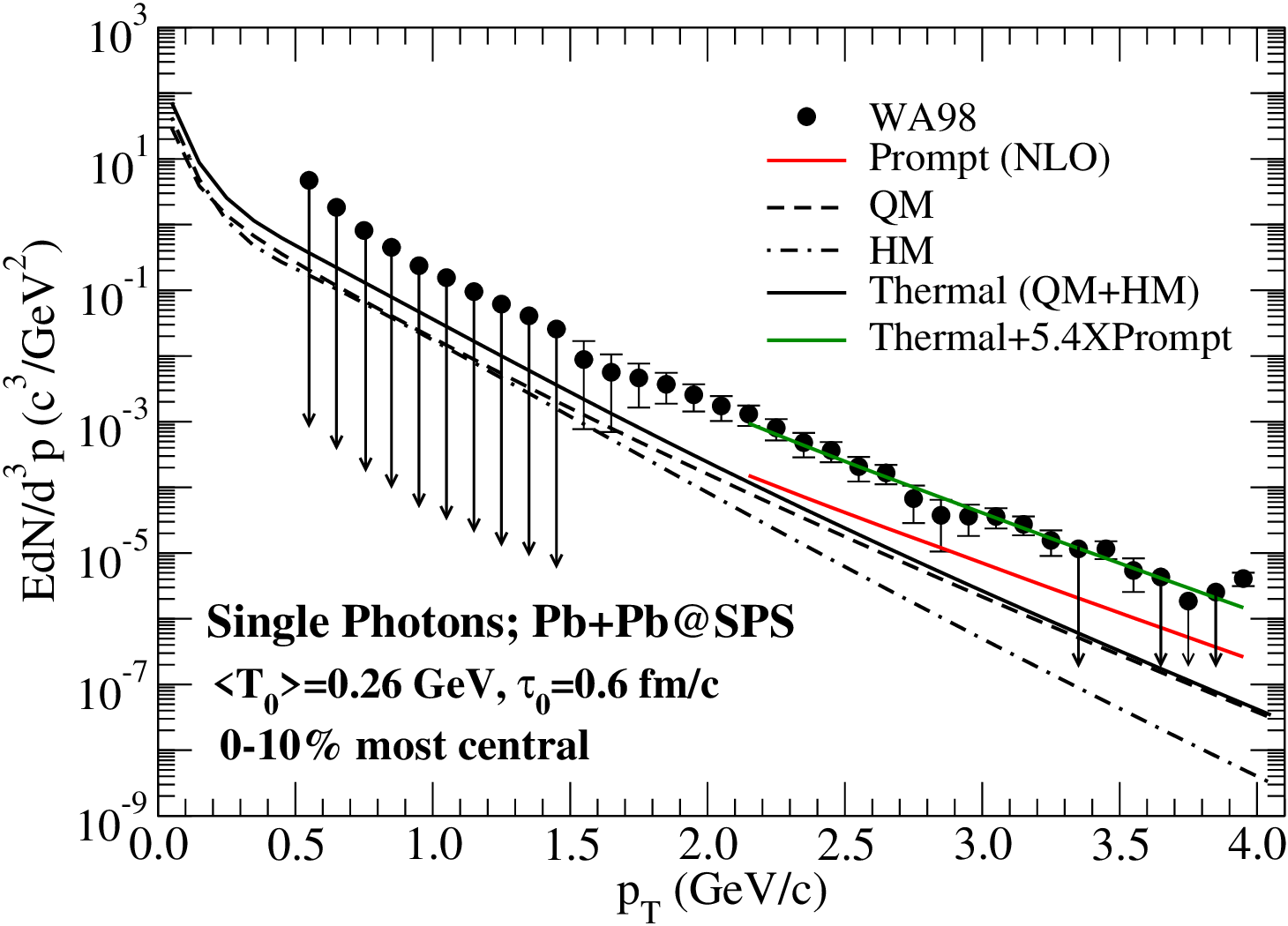}}
\centerline{\includegraphics*[width=7.5cm]{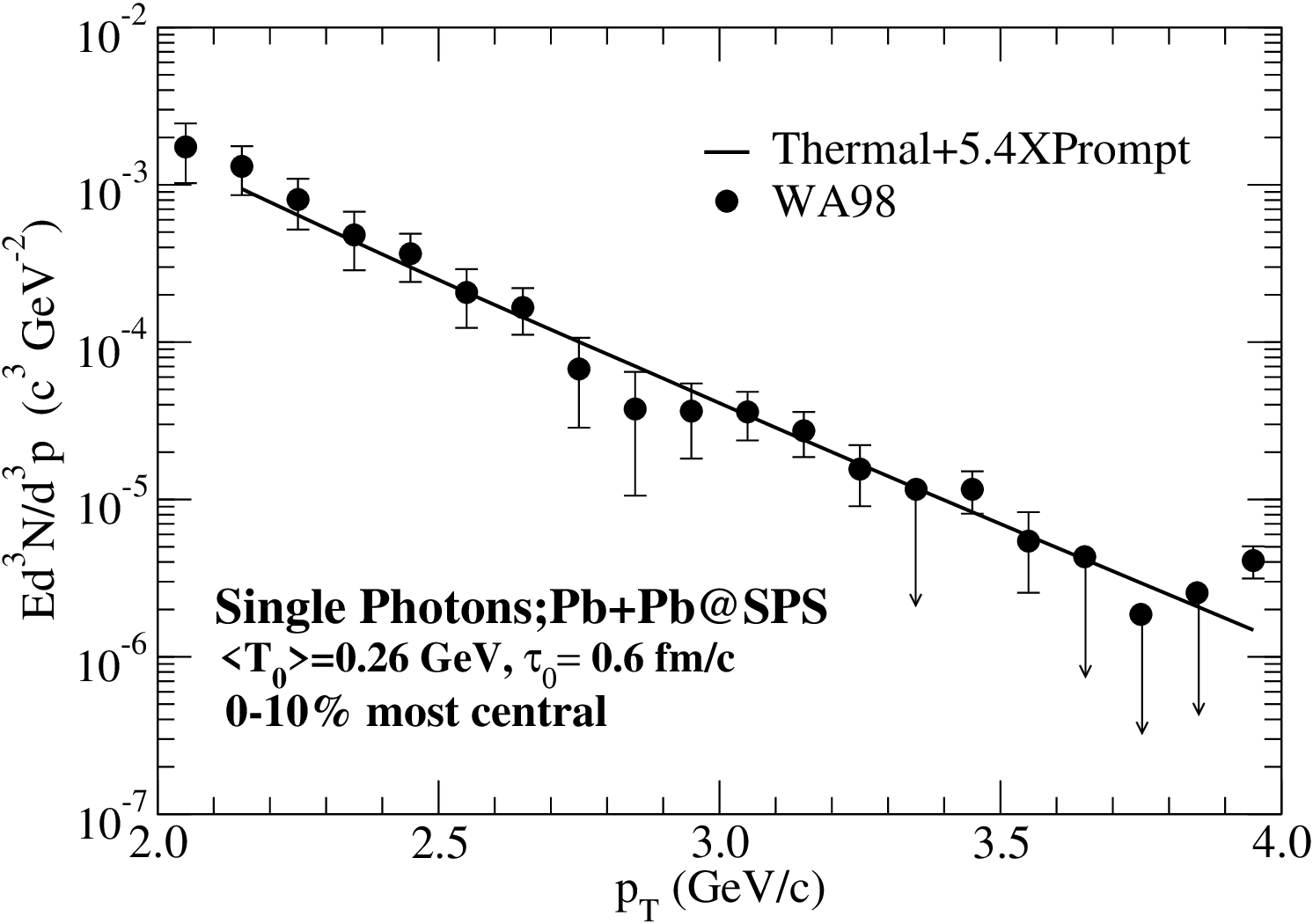}}
\centerline{\includegraphics*[width=7.5cm]{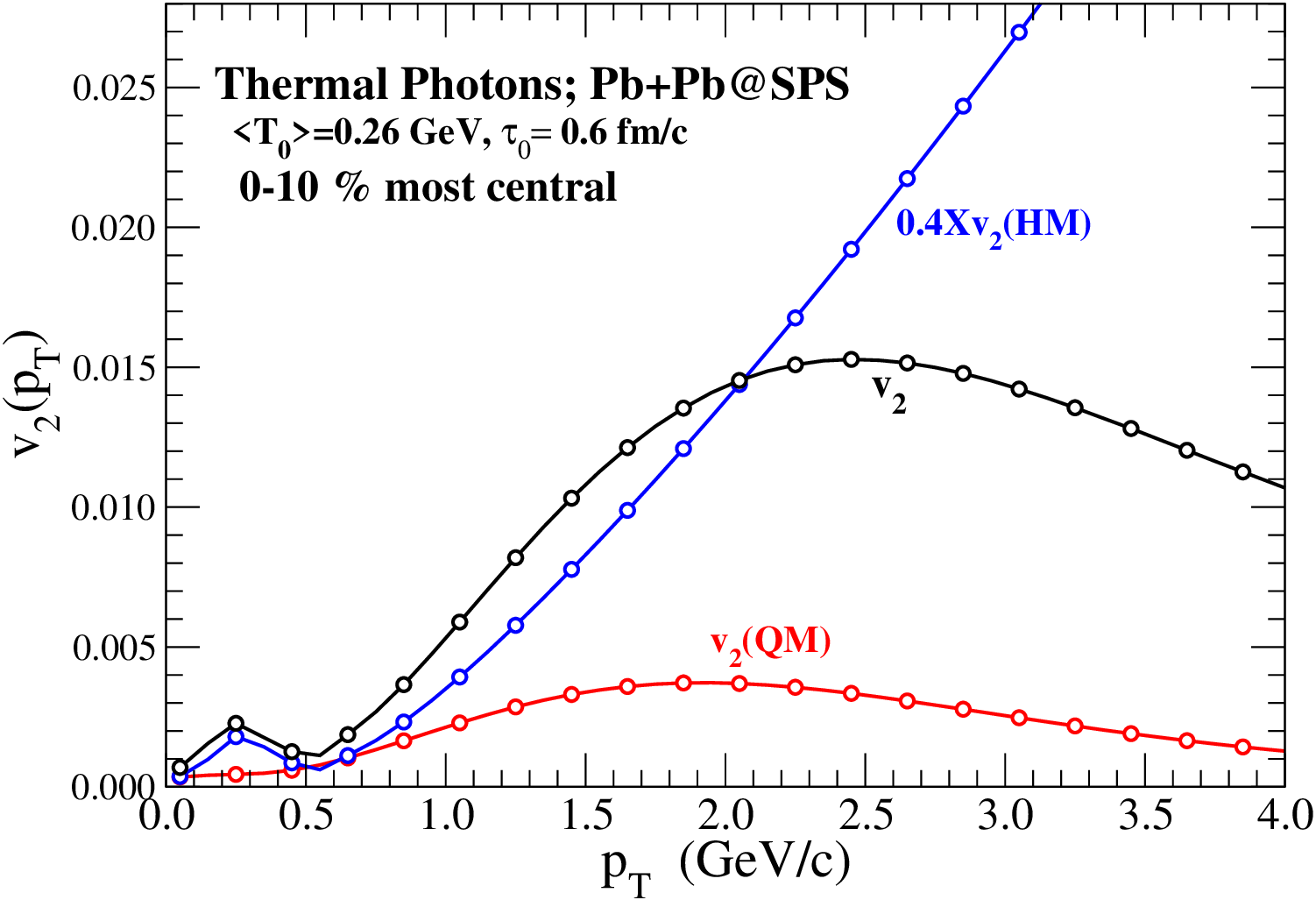}}
\caption{Same as Fig.~\ref{fig5} for $\tau_0$ = 0.6 fm/$c$. 
Note that the NLO pQCD results have to be scaled up by 5.4 
for describing the data} 
\label{fig7}
\end{figure}

\begin{figure}[ht]
\centerline{\includegraphics*[width=7.5cm]{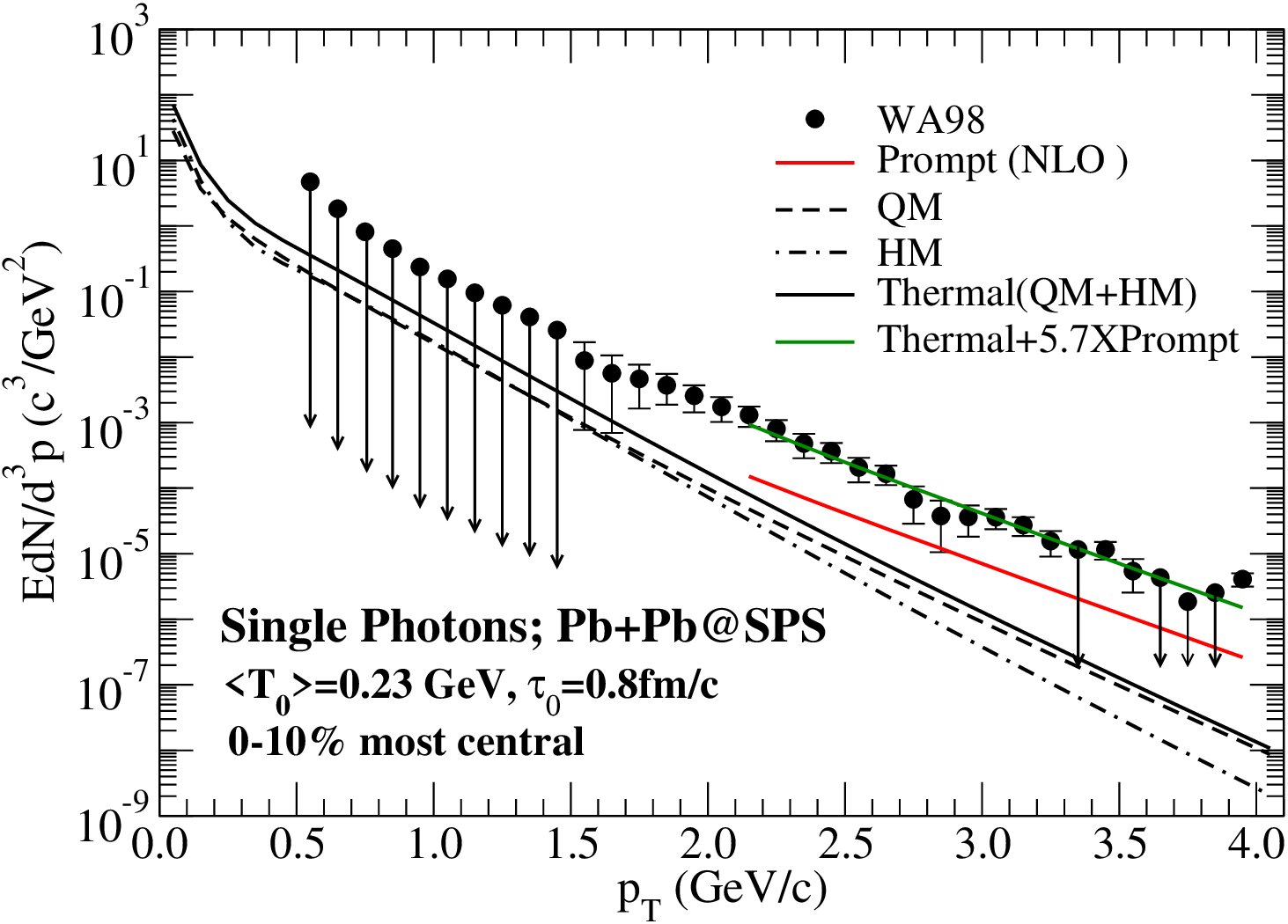}}
\centerline{\includegraphics*[width=7.5cm]{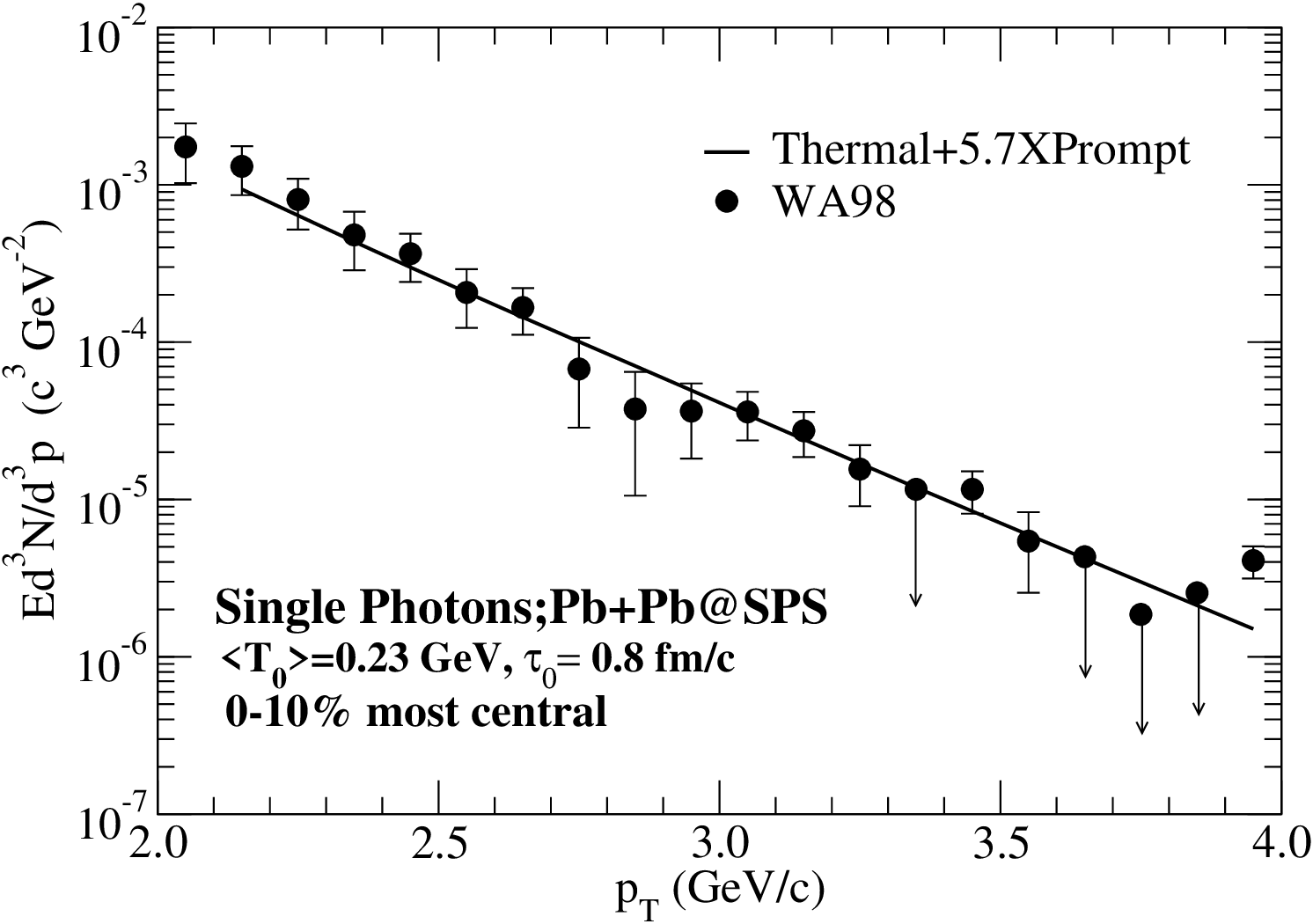}}
\centerline{\includegraphics*[width=7.5cm]{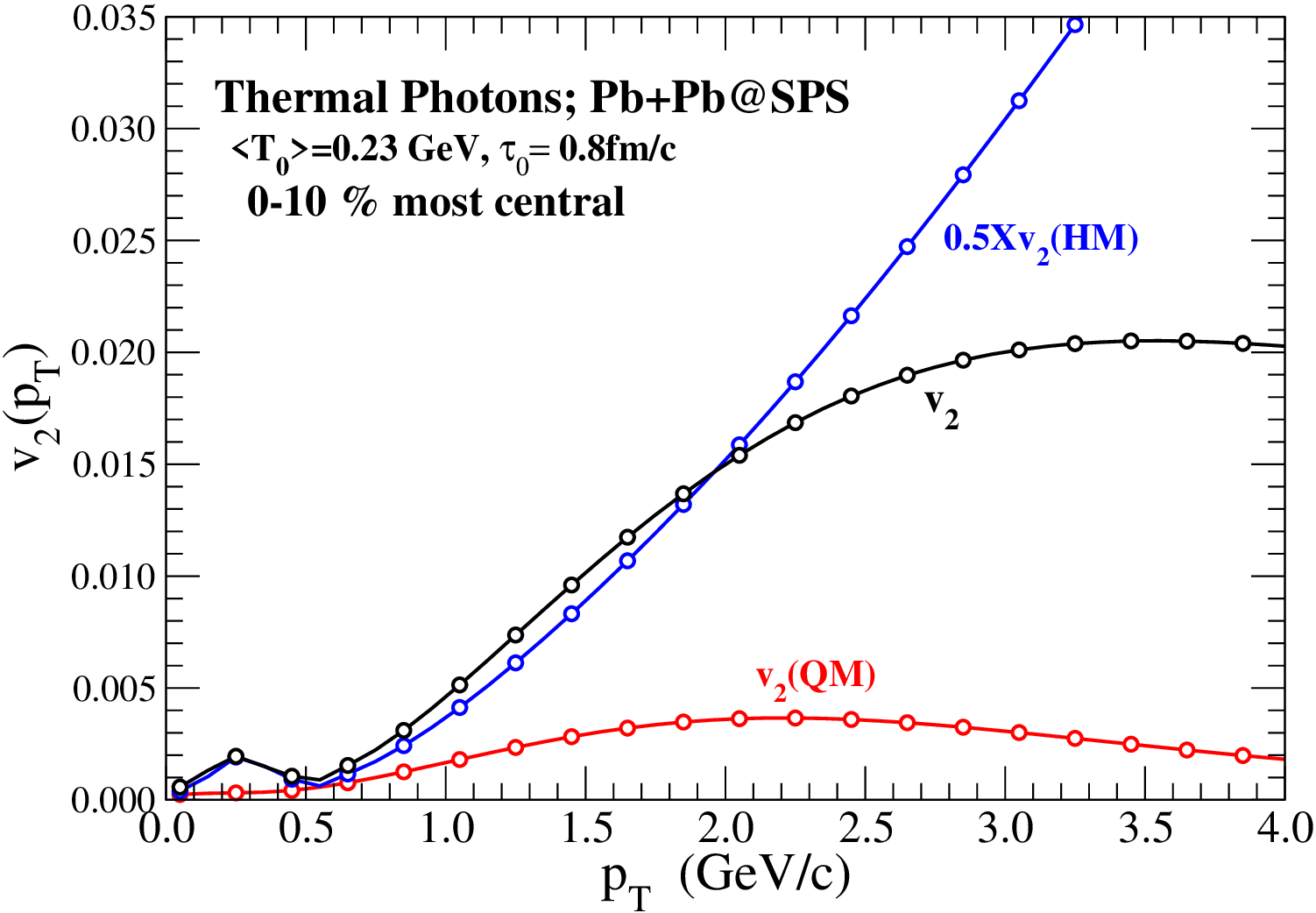}}
\caption{Same as Fig.~\ref{fig5} for $\tau_0$ = 0.8 fm/$c$. 
Note that the NLO pQCD results have to be scaled up by 5.7 
for describing the data} 
\label{fig8}
\end{figure}

\begin{figure}[ht]
\centerline{\includegraphics*[width=7.5cm]{pb_1.0.eps}}
\centerline{\includegraphics*[width=7.5cm]{1.0.eps}}
\centerline{\includegraphics*[width=7.5cm]{1.0_v2.eps}}
\caption{Same as Fig.~\ref{fig5} for $\tau_0$ = 1.0 fm/$c$. 
Note that the NLO pQCD results have to be scaled up by 5.9 
for describing the data} 
\label{fig9}
\end{figure}

\begin{figure}[ht]
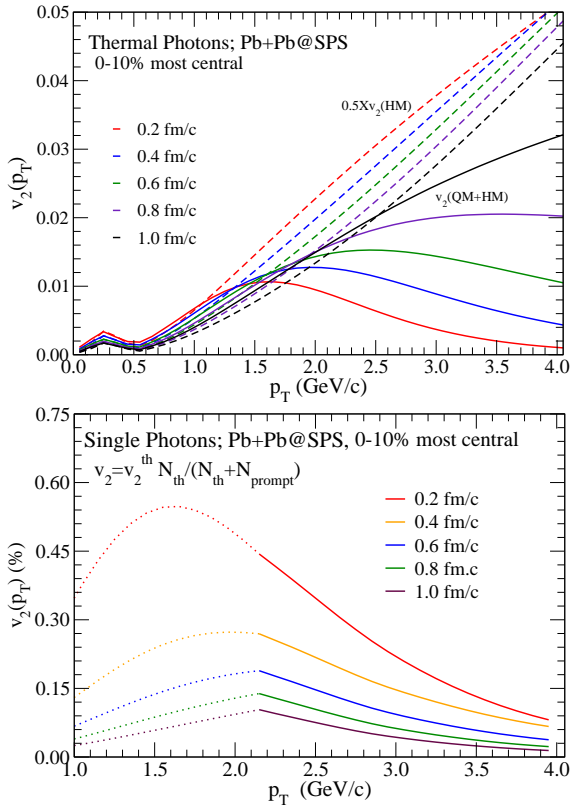

\centerline{\includegraphics*[width=7.5cm]{all_v2.eps}}
\centerline{\includegraphics*[width=7.5cm]{frac.eps}}
\caption{Upper Panel: $v_2$ for thermal photons for different 
$\tau_0$. The results for photons from hadronic matter alone  
are also given. Lower Panel: $v_2$ for single photons. Results 
below $p_T$ = 2.55 GeV/$c$ (shown as dotted curve) are obtained 
by arbitrarily using the normalizing factor at $p_T$ = 2.55 GeV/$c$. 
It is assumed that the difference of experimental data and the 
thermal production can be attributed to prompt and pre-equilibrium 
contributions.}
\label{fig10}
\end{figure}

\begin{figure}
\centerline{\includegraphics*[width=7.5cm]{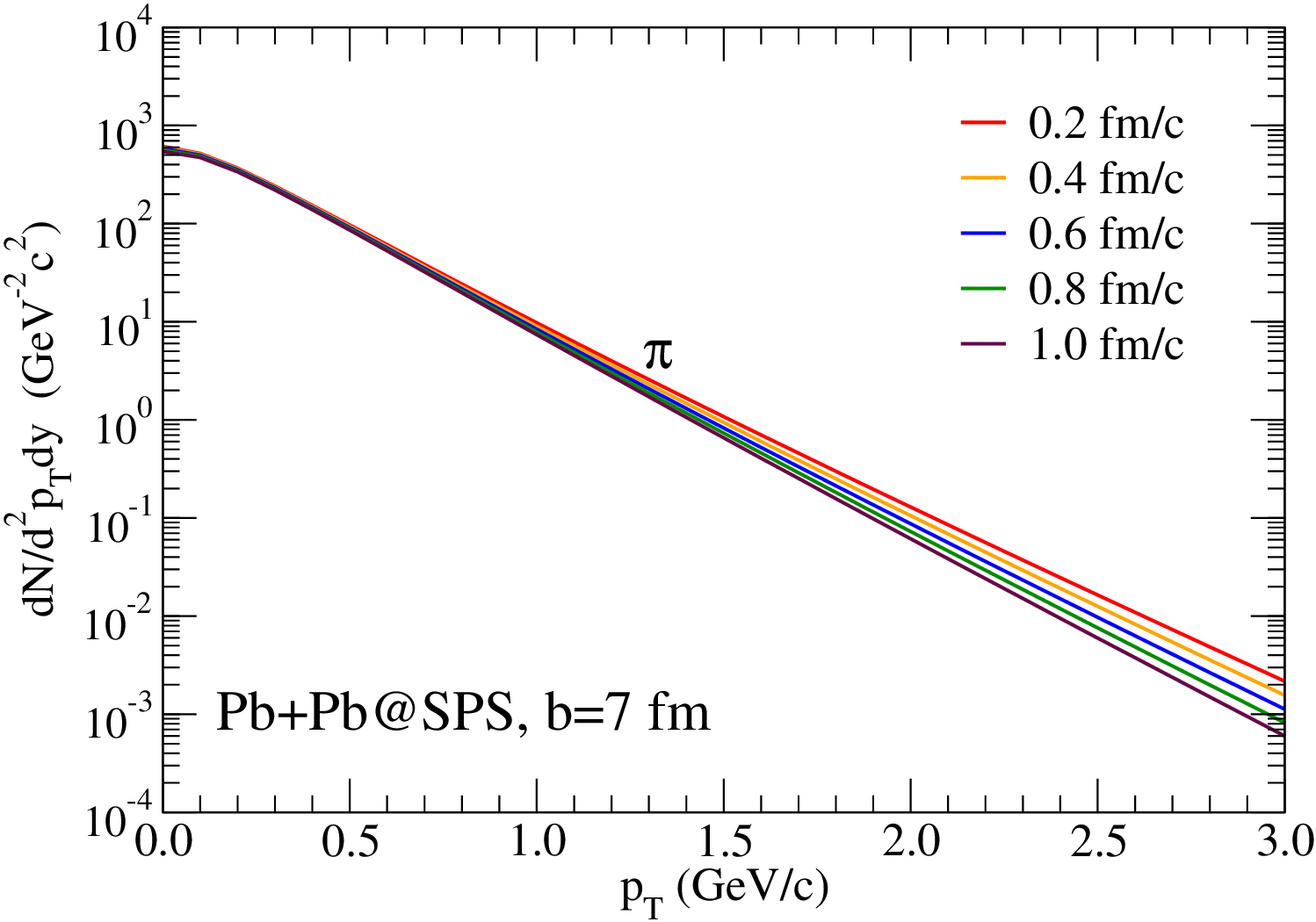}}
\centerline{\includegraphics*[width=7.5cm]{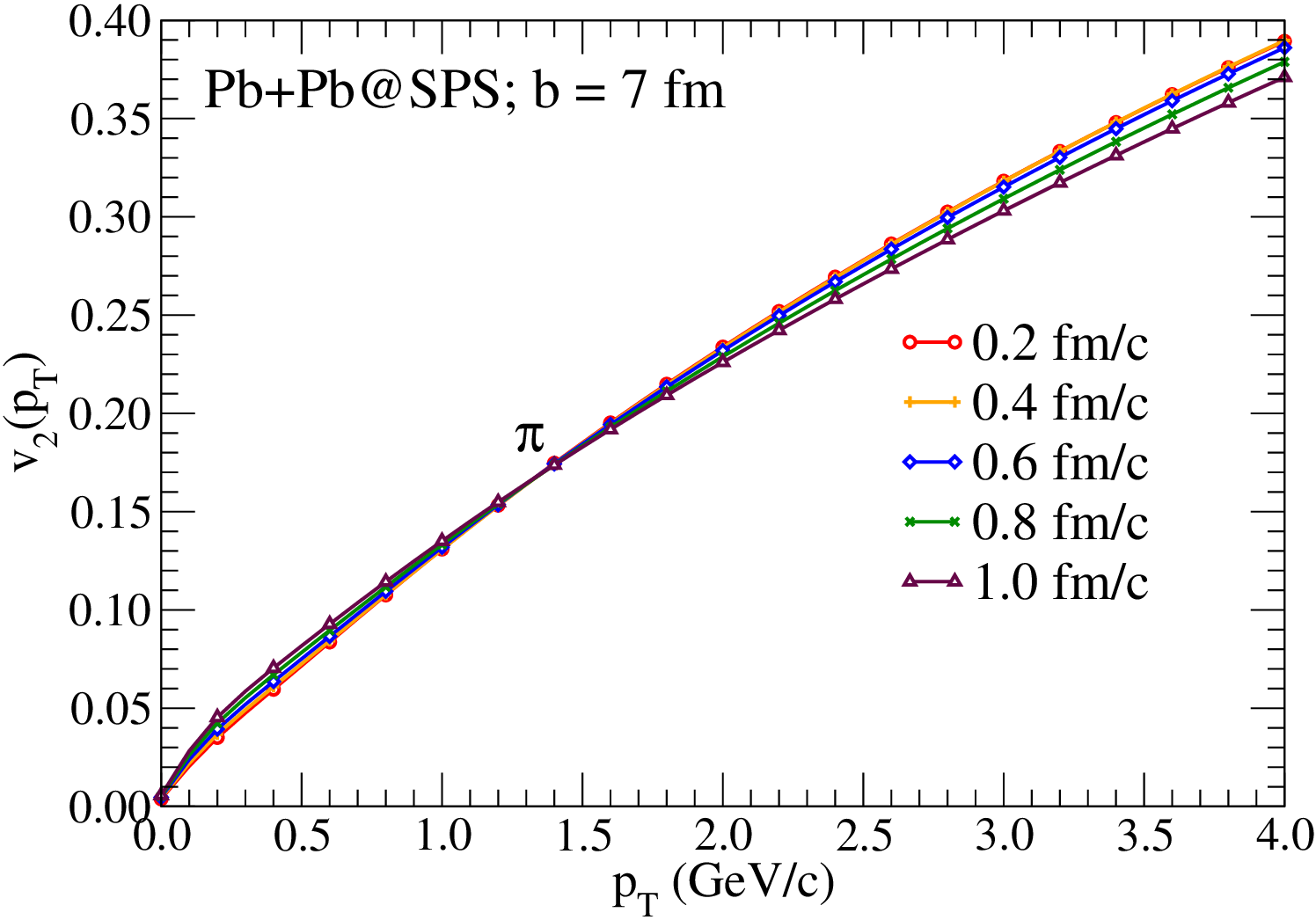}}
\caption{Spectra (upper pannel) and $v_2(p_T)$ (lower panel) '
for primary pions from $Pb+Pb$ collisions having $b=$ 7 fm at 
SPS energy, for different initial times.}
\label{fig11}
\end{figure}

\begin{figure}
\centerline{\includegraphics*[width=7.5cm]{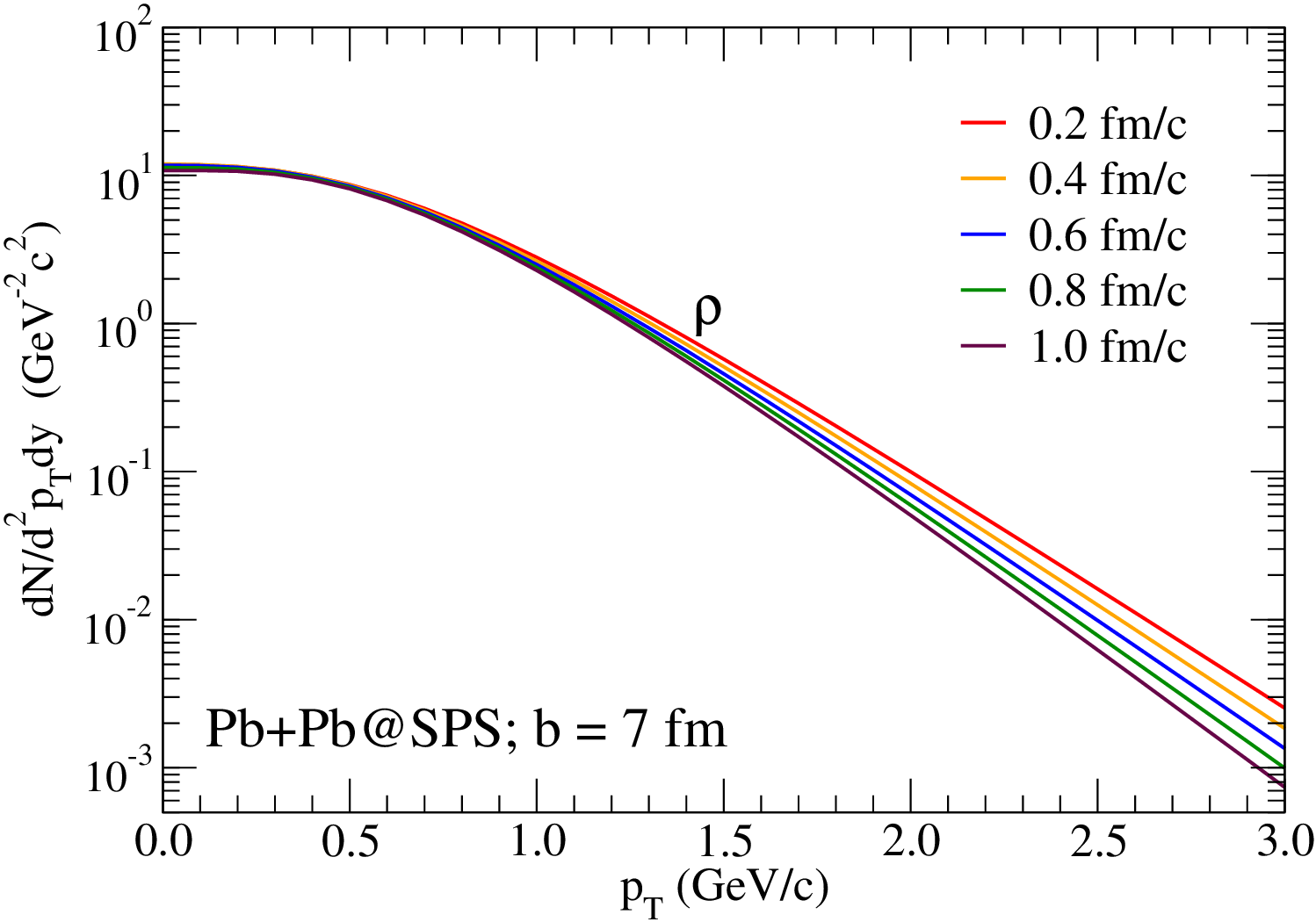}}
\centerline{\includegraphics*[width=7.5cm]{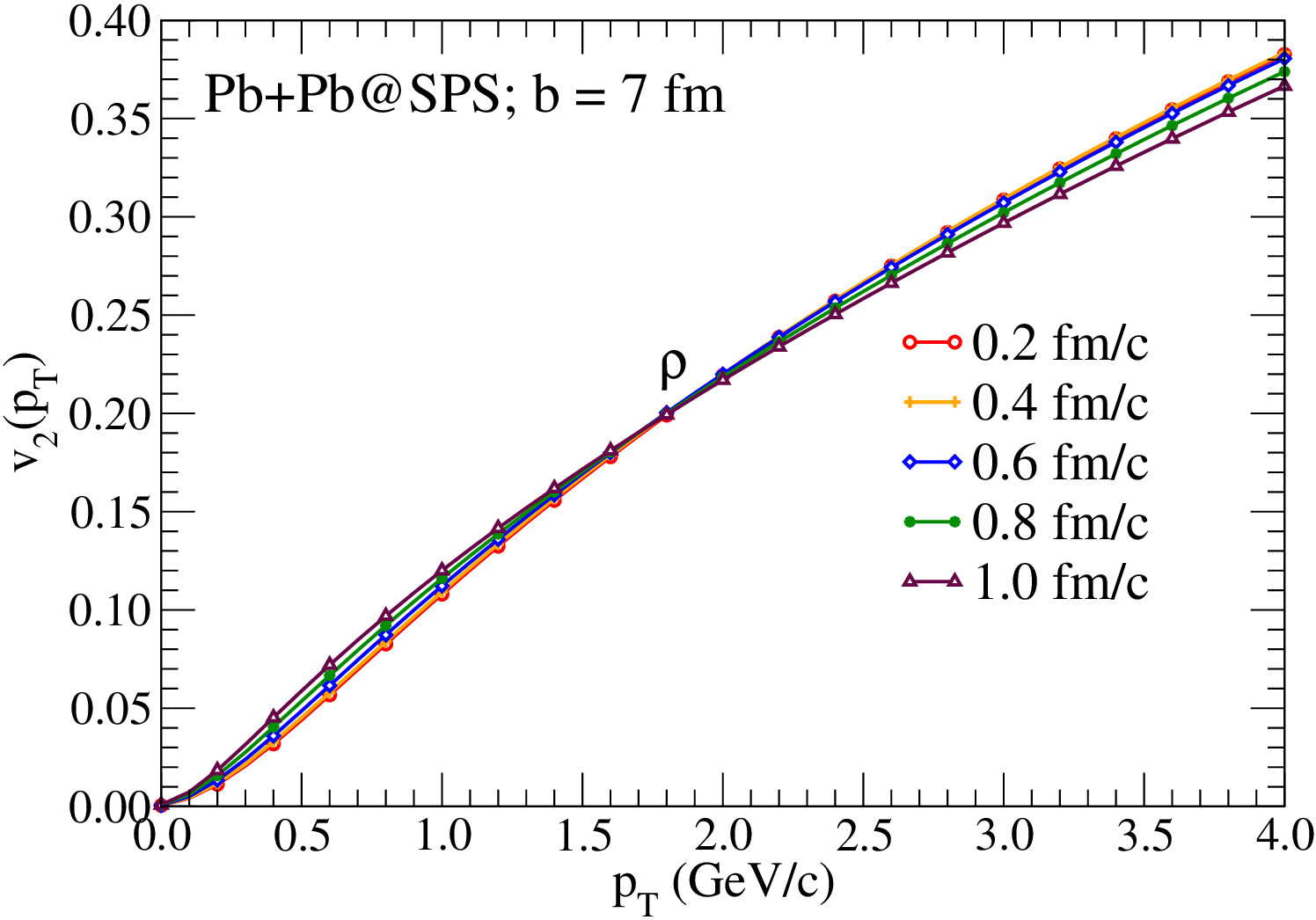}}
\caption{Spectra (upper panel) and $v_2(p_T)$ (lower panel) 
for primary rho-mesons from $Pb+Pb$ collisions having $b=$ 7 fm 
at SPS energy, for different initial times.}
\label{fig12}
\end{figure}

\begin{figure}
\centerline{\includegraphics*[width=7.5cm]{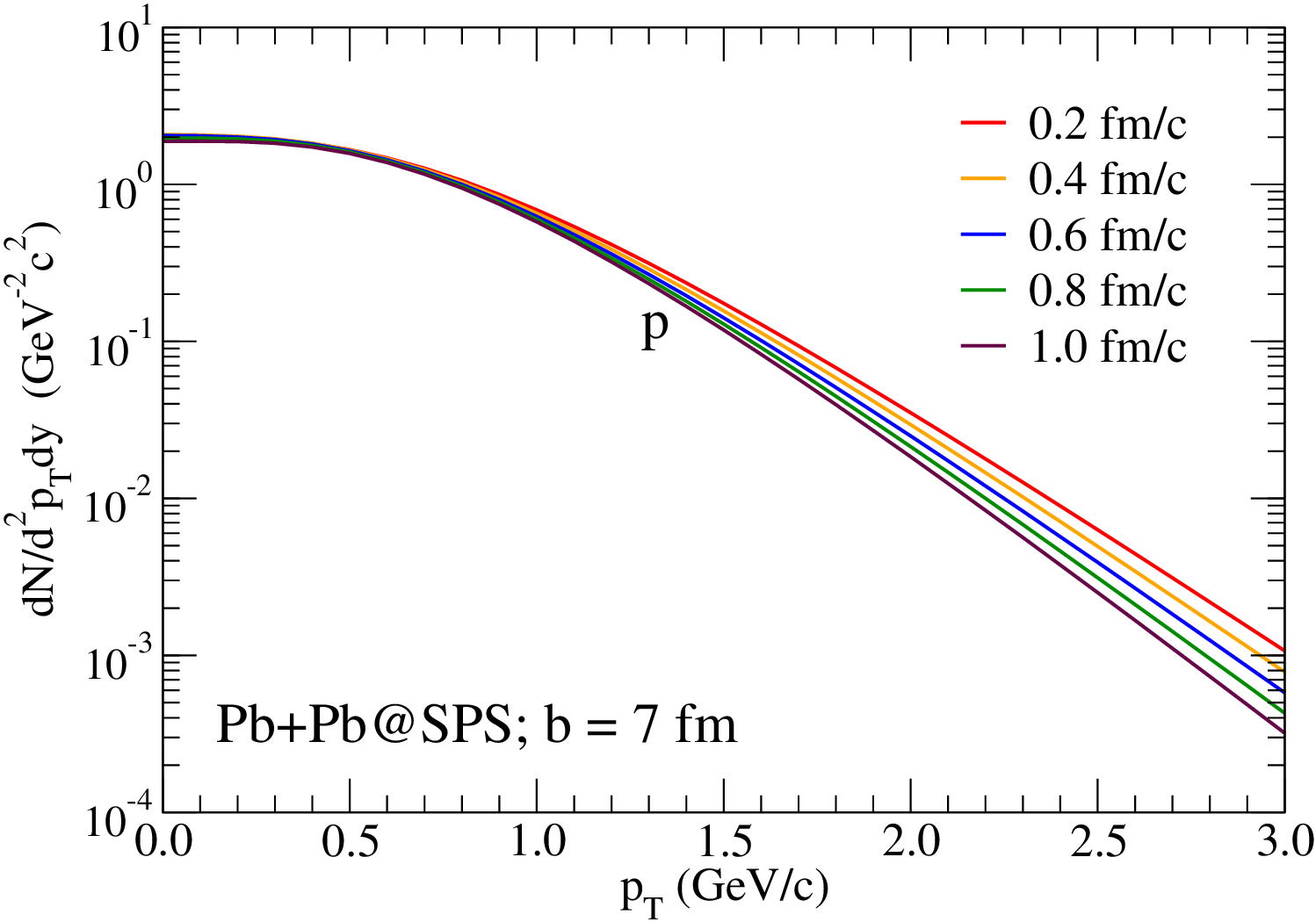}}
\centerline{\includegraphics*[width=7.5cm]{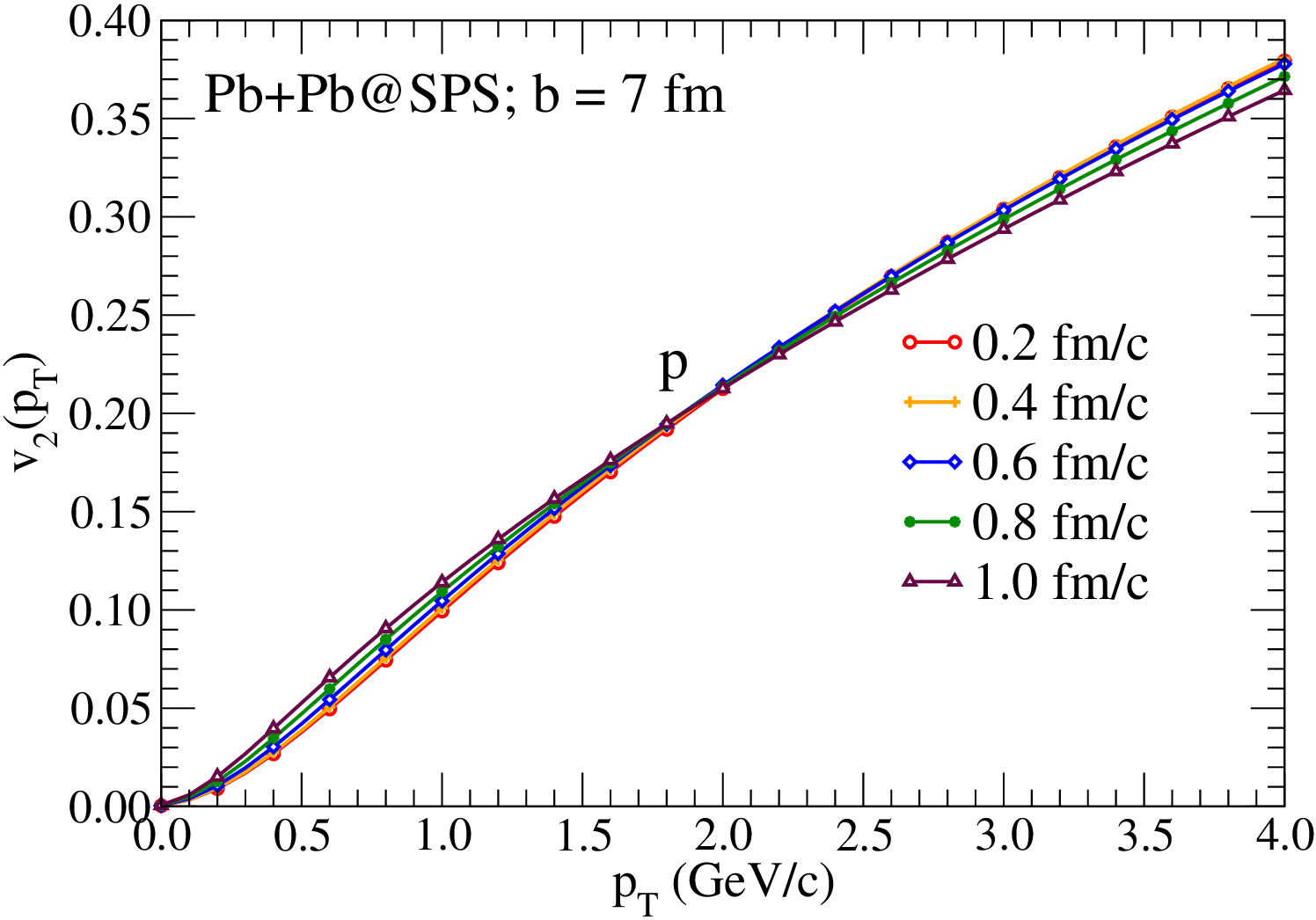}}
\caption{Spectra (upper panel) and $v_2(p_T)$ (lower panel) 
for primary protons from $Pb+Pb$ collisions having $b=$ 7 fm 
at SPS energy, for different initial times.}
\label{fig13}
\end{figure}

\section{Prompt Photons}

As mentioned earlier, it is  quite crucial to get an accurate 
estimate of prompt photon production in nucleus-nucleus collisions, 
before we can 
%
% SY's mod
%embark on the journey -> begin
%
begin to explore the initial conditions of the thermalized system.
The success of PHENIX experiment in measuring the single photon 
production at large $p_T$ in $Au+Au$ collisions at RHIC energies has 
brought this consideration into a sharp focus. Thus, for example, 
it is now realized~\cite{gale1} that the `suppression' of single 
photons at large $p_T$ in  $Au+Au$ collisions compared to those 
from $pp$ collisions at the same nucleon-nucleon centre of mass 
energy has its origin predominantly in the difference of isospin 
for protons and neutrons (or their valence quark structure).

This is often overlooked when the $pp$ data are scaled by the 
nuclear thickness $T_{AA}(b)$, for the above comparison.  Of 
course, one additionally needs to account for effect of 
%
%quenching of jets, -> jet energy loss
%
jet energy loss if a quark-gluon plasma is formed. The study 
of prompt photon production in $pp$ collisions has reached a 
high degree of sophistication. All the available data have now 
been analyzed using NLO pQCD~\cite{pat1,pat2,vogel1,vogel2} and 
it is generally believed that choosing the factorization, 
renormalization, and fragmentation scales as equal to $p_T$/2 
provides an excellent description to all the single photon data 
{\em except} for those from  the E704~\cite{e704} and the 
E706~\cite{e706} experiments, without the requirement of any 
intrinsic $k_T$.  Inclusion of intrinsic $k_T$ improves 
the description of the these data but simultaneously destroys 
the good agreement with all the other data. The E704 data is at 
$\sqrt{s}=$ 19.4 GeV, which is close to the $\sqrt{s_{NN}}\approx$ 
17.3 GeV relevant for the WA98 experiment. Two other experiments 
NA3~\cite{na3} and E629~\cite{e629} have measured single photon 
production from $p+^{12}C$ collisions at the same energy and these 
data are often used with a normalization by the mass-number of the 
target to estimate the $pp$ data, even though half of the nucleons 
in the target are neutrons. We have verified that accounting for 
this reduces the 
%
%SY:
%theoretical prediction -> theory values
%
%
theory values by about 2\% at $p_T\approx$ 2 GeV/$c$ and 
by about 15\% for $p_T\approx$ 6 GeV/$c$, which is well below 
the other experimental uncertainties. We shall ignore this for 
the moment. We show our
%
%prediction -> calculation
%
calculation for the single photon production 
at $\sqrt{s}$= 19.4 GeV in Fig.~\ref{fig1}, along with the data 
reported by the NA3~\cite{na3},~\cite{e704}, and E629~\cite{e629} 
experiments. We note that the fragmentation contribution at this 
energy is of the order of 30\% of the Compton + annihilation term. 
We  also see that the NLO pQCD provides a good description of the 
NA3 data, while it underestimates the E704 and E629 data by a factor 
of 2--6. This has been noted by several studies as mentioned 
earlier~\cite{pat1,vogel2} and it is known that these data deviate 
also from the $x_T$ scaling which all the other data 
follow~\cite{scale_phot}. We have verified that this scaling 
is in good agreement with the NLO pQCD results for values of 
$p_T$ up to about 4.5 GeV/$c$, but over-predicts the results 
considerablyat higher $p_T$. We are discussing this point again 
as several studies have tried to accommodate these data by 
incorporating intrinsic $k_T$ for the partons, which is not 
favoured by the rest of the data. We must add though these 
results are among the earliest measurements of single photons,
which may account for the inconsistency of data between different 
experiments and even within the same experiment~\cite{na3}.

Recently, prompt photon production in $p+^{12}C$ and $p+^{208}Pb$ 
collisions at CERN SPS energy appropriate for the WA98 
experiment~\cite{baumann} has been measured. Only the upper 
limit of the single photon production could be deduced. We have
verified that the upper limits are about a factor of 5--10 larger
than the NLO pQCD calculations, though the slope of the data is 
described well by the calculations.

We also note that the inclusion of intrinsic $k_T$ of partons 
is not easy even at lowest order of pQCD~\cite{wong_kt,dum_kt} 
and the results for NLO pQCD are often inferred by using a K-factor  
which describes the difference of results of lowest order pQCD 
with and without the intrinsic $k_T$ (see also Ref.~\cite{nlo_kt}). 
In a nuclear medium, Cronin effect also contributes to the broadening 
of the transverse momentum spectrum. In Ref.~\cite{simon}, it was 
reported that this broadening by the Cronin effect can lead to
an enhancement of photon production by a factor of about 2.5 
for  $p_T$ of 2--4 GeV/$c$ for the case of the WA98 experiment.
%
%An interesting approach for this was 
%adopted by Turbide et al~\cite{simon}, who estimated the effects of 
%the  by folding an empirically fitted distribution for the 
%experimental data for $pp$ collisions with an additional Gaussian 
%distribution for the transverse $k_T$ for partons for $AA$ collisions.
%

Our task of obtaining an accurate estimate of prompt photon 
production in $Pb+Pb$ collisions at $\sqrt{s_{NN}}=$ 17.3 GeV 
is further complicated. Firstly, there is no single photon 
production data from $pp$ collisions at this energy. In any 
case there is no data for $pn$ and $nn$ collisions, which will 
also contribute to the production of prompt photons from lead 
nuclei. The importance of these can be seen from Fig.~\ref{fig2}
where we show our results for the effect of isospin and shadowing
for prompt photon production at this energy. We see that effects 
of isospin and shadowing reduce the single photon production at  
$p_T=$ 4 GeV/$c$ by about 30\% for $Pb+Pb$ collisions compared to 
naive scaling of $pp$ data by the nuclear thickness used in most 
of the early studies, including those involving one of the present 
authors~\cite{dks}. We also note that the inclusion of shadowing 
leads to a 
%
%SY's mod
%
%
%richness in the data, -> siginficant variation in the end result
%
significant variation in the end result so that at lower 
transverse momenta, the single photon production goes up.
The middle and the lower panels of the figure show the 
results for $x_T=2p_T/\sqrt{s}$ scaling with the inclusion 
of iso-spin and parton shadowing at SPS, RHIC, and LHC 
energies\cite{emprobes}. The deviations from a scaling
behaviour are due to the scale dependence of the structure 
functions and the QCD interactions~\cite{owens}.

In the light of the discussions above we take the following 
view for getting the yield of prompt photons for $Pb+Pb$ 
collisions at an energy corresponding to the WA98 experiment. 
For a given impact parameter $b$, we first estimate the 
effective number of protons and neutrons from the number 
of participants:
\begin{eqnarray}
N_{\text{part}}(b)= \int  \,dx \, dy \, \nu(x,y,b)
\end{eqnarray}
where
\begin{eqnarray}
&&\nu(x,y,b)
=
\nonumber\\
&&
\Big\{ 
T_A(x{+}{b}/{2},y)
\left[1-\left(1-
{\sigma T_B(x{-}{b}/{2},y)}/{B}\right)^B\right]\nonumber\\
&&+T_B(x{-}{b}/{2},y)
%\times\nonumber\\ & &
\left[1-\left(1-
{\sigma T_A(x{+}{b}/{2},y)}/{A}\right)^A\right]\Big\}\, .
\label{eq:nuxyb}
\end{eqnarray}
is the surface density.
In the above $T_A$ is the nuclear thickness function
of the nucleus A;
\begin{equation}
T_A(x,y)=\int_{-\infty}^{+\infty} dz \, \rho_A(x,y,z)\, ,
\end{equation}
where the nuclear density is given by a Woods-Saxon distribution,
\begin{equation}
\rho_A(r)=\frac{\rho_0}{1+\exp\left[(r-R)/a\right]} \, ,
\end{equation}
with the normalization,
\begin{equation}
\int d^3r \, \rho_A(r)=A \, .
\end{equation}
A similar expression holds for the nucleus B. We shall take 
the nuclear radius $R$ for the Pb nucleus to be 6.5 fm and 
the diffuseness $a$ to be 0.54 fm. The nucleon-nucleon 
inelastic cross-section $\sigma$ is set to 32 mb relevant 
for the $\sqrt{s_{NN}}=$ 17.3 GeV.

Now we assume that the effective number of protons from the 
projectile or the target which constitute the participants 
at a given impact parameter $b$ is given by;
\begin{equation}
Z^{\text{eff}}_{\text{Proj}}=Z^{\text{eff}}_{\text{Targ}}=
\frac{Z}{A}\frac{N_\text{part}(b)}{2}\, ,
\end{equation}
with a similar expression for the effective number of neutrons. 
These numbers then decide the shadowing functions $R_{A_{\text{eff}}}
(x,Q^2)$ as well as the effective structure functions,
\begin{equation}
f_{A_{\text{eff}}}(x)=\frac{Z_{\text{eff}}}{A_{\text{eff}}} f_p(x) +
\frac{N_{\text{eff}}}{A_{\text{eff}}} f_n(x)
\end{equation}
for the prompt photon calculations. We multiply the cross-sections
for the production of photons from the  ``effective'' nucleon-nucleon 
collisions using NLO pQCD, with the nuclear overlap function;
\begin{equation} 
T_{AB}(b)=\int dx \, dy \, T_A(x+\frac{b}{2},y)T_B(x-\frac{b}{2},y)
\end{equation}
to get the yield of prompt photons. Finally the yield is 
averaged over the impact-parameter range covered by the 
centrality of the collision.

\section{Thermal Photons}
\subsection{Initial conditions}

We have already noted that the importance of thermal photons 
lies in their sensitivity to initial conditions. The simplest 
and most widely used initial conditions assume formation of 
a hot, thermalized, and chemically equilibrated quark gluon 
plasma at some initial time $\tau_0$, beyond which the system
expands isentropically ignoring the viscosity effect. This makes
the powerful methods of hydrodynamics available to us. One may 
also use a parametrized fire-ball to describe the evolution 
of the system. 

%One could use the parton cascade model~\cite{kkg,bms} to model the evolution 
%of the system starting from the initial nuclei and treating the scattering, 
%radiation, and fusion of hard partons within pQCD.
%This, however, precludes the
%participation of the bulk of the soft partons 
%which, by definition, cannot contribute to hard scatterings.
%The soft part, however, can form strong colour fields that can influence
%the propagation of hard partons through instabilities. 
%The parton cascade model, as well as, minijet based  models~\cite{biro,eskola}
%do not produce
%a thermally and chemically equilibrated plasma. One can however 
%assume the thermalization of the produced partons and then devise master 
%equations to study the evolving chemical equilibration\cite{biro,munshi,sspc,
%gelis}. 
%The hadronization of a chemically non-equilibrated plasma, 
%however, raises 
%questions about the bag-constant and transition temperature, which 
%depend on quark and gluon fugacities if the usual Gibb's criterion of
%equality of pressure at the transition temperature are used (see Ref.
%~\cite{gelis} for a possible solution). This has not been attempted for 
%energies relevant for SPS, as the number of partons produced when a 
%realistic $p_T$ cut-off is used is too low.   
%Even though, the parton cascade model excludes the underlying soft-sector 
%it may still
%be used with advantage to get photon production at larger $p_T$ due to 
%multiple scatterings and radiations~\cite{bms_phot1,bms_phot2}. 

For this study, we employ a boost invariant hydrodynamics \cite{ksh} 
as our model of the underlying bulk evolution, especially for the 
purpose of obtaining the initial energy and temperature distributions.
This model has been used extensively to explore and hadron production 
and elliptic flow of hadrons as well as photons~\cite{v2_phot} and 
dileptons~\cite{v2_dil}. For the SPS energies under consideration, 
the initial conditions are estimated by assuming~\cite{ksh,ss_sau,
ss_80,ss_98} that the deposited energy in the transverse plane is 
proportional to the number of wounded-nucleons~\cite{wa80_Et} 
(or participants);
\begin{eqnarray}
\epsilon(x,y,b,\tau_0)
&=& K\,\nu(x,y,b)
%\left[ \,T_A(x+\frac{b}{2},y)\right. \nonumber\\
%&\times& \left[1-\left(1-
%\frac{\sigma T_B(x-\frac{b}{2},y)}{B}\right)^B\right]\nonumber\\
%&+&T_B(x-\frac{b}{2},y)\times\nonumber\\
%& &\left[1-\left(1-
%\frac{\sigma T_A(x+\frac{b}{2},y)}{A}\right)^A\right]\, .
\end{eqnarray}
where $K$ is a constant and $\nu(x,y,b)$ is given in Eq.(\ref{eq:nuxyb}).
We further assume, as in Ref.~\cite{ksh} that the initial transverse 
density profile of net baryon number is proportional to the 
participant profile as well
\begin{equation}
n(x,y,b,\tau_0)=L \, \epsilon(x,y,b,\tau_0)\, .
\end{equation} 

The authors of Ref.~\cite{ksh} have shown that taking $\tau_0$ 
= 0.8 fm/$c$ along with $K$ = 2.04 GeV/fm and $L$ = 0.122 
GeV$^{-1}$ provides a remarkably quantitative description of 
the particle spectra measured for the $Pb+Pb$ collisions at 
$\sqrt{s}_{NN}=17.3$\,GeV, using the equation of state Q, 
which provides that the thermally and chemically equilibrated 
QGP undergoes a first order phase transition to hadrons at 
$T_c \approx$ 164 MeV. We have checked that these values give 
a quantitative description of the deposited transverse energy 
measured by the WA98 experiment~\cite{wa98_Et} for central 
collisions. 

In the  present work we use these values for $K$, $L$, and 
$\tau_0$. We additionally explore the consequences of varying 
$\tau_0$ such that  the rapidity density of total entropy, 
$dS/dy$, and net baryons, $dN_{B\overline{B}}/dy$, remains 
fixed (see Ref.\cite{ss_98,pasi} for a similar approach).
This is attained by taking the entropy density $\propto 
\epsilon^{3/4}$ and then using $s_0\tau_0$ and $n_0\tau_0$ 
as constants, where $s_0=s(x=0,y=0,b=0)$ and $n_0=n(x=0,y=0,b=0)$.
This corresponds to an isentropic expansion. Note that the shape 
of the these distributions are taken as independent of $\tau_0$.
Thus we have used $\tau_0=$ 0.2, 0.4, 0.6, 0.8, and 1.0 fm/$c$. 
These then correspond to the peak temperature, $T_0(x=0,y=0,b=0)$ 
of 420, 330, 284, 257, and 238 MeV respectively, while the average 
temperatures are 380, 295, 257, 233, and 215 MeV, respectively.

\subsection{The Flow Patterns}

As a first step we determine the time-evolution of the average 
energy density, average temperature, and the average transverse
velocity of the expanding system obtained from the hydrodynamic
calculations for a central collision. We take the average by 
defining,
\begin{equation}
\left< f \right > =\frac{\int  \, dx \, dy \, f(x,y) \, \epsilon(x,y,\tau)}
{\int \, dx \, dy \, \epsilon(x,y,\tau)}
\end{equation}
 We note (see Fig.~\ref{fig3}) that in the over-lapping 
time-span, the variations of these quantities are quite similar, 
though an earlier start leads to  a slightly larger build-up of 
the flow velocity and a faster cooling of the system. In the 
final stages the temperatures and the velocities do not differ 
beyond about 10\% for different initial times, though the 
energy density varies by about 40\% (as it varies as $T^4$). 

Photons are sensitive to the initial temperature. Therefore, an 
earlier initial time with higher initial temperature will lead 
to a considerably enhanced production of photons at higher 
transverse momenta. On the other hand, this should not affect 
the spectra of hadrons since they are emitted at much later 
freeze-out stage when the effect of having different initial 
times is mostly washed out.

In Fig.~\ref{fig4} we have shown the time-evolution of the average 
effective temperature (or the blue-shifted temperature) to see the
combined effect of the cooling and expansion (velocity). We define 
the effective temperature as,
\begin{equation}
T_{\text {eff}}=T \, \sqrt{\frac{1+v_T}{1-v_T}} \, .
\end{equation}
We note that as in Fig.~\ref{fig3}, the results differ only marginally
beyond the time of about 1 fm/$c$, confirming once again our surmise
that the difference in the production of thermal photons should mostly 
arise from contributions before this time.

\subsection{Thermal photons}

We calculate the production of thermal photons by folding the history of
the evolution of the system with the rate for the production of photons
from the quark matter and the hadronic matter. We use the complete 
leading-order results for the production of photons from the QGP from 
Arnold, Moore, and Yaffe~\cite{guy} and the latest results for the 
radiation of photons from a hot hadronic gas obtained by Turbide, Rapp, 
and Gale~\cite{simon}. As mentioned earlier, the equation of state 
(EOS Q~\cite{ksh}) incorporating a phase transition to quark gluon 
plasma at $T\approx$ 164 MeV, and resonance gas for the hadronic phase 
below the energy density of 0.45 GeV/fm$^{-3}$ is used to describe the 
evolution. The mixed phase is described using Maxwell's construction.  
The freeze-out is assumed to take place at $\epsilon=$ 0.075 GeV/fm$^3$. 
Final results are obtained by taking an average of the results over the 
range of impact parameters $b$ between 0 and 4.6 fm corresponding to 
0--10\% most central collisions, considered by the WA98 experiment. 

We summarize our results for the case of $\tau_0$ =0.2, 0.4, 0.6, 0.8, and 
1.0 fm/$c$ in Figs.~\ref{fig5},\ref{fig6},\ref{fig7},\ref{fig8},\ref{fig9}.
As expected from the discussion earlier, we find that the hadronic matter 
contribution to the single photons is only marginally altered as we increase
the initial time or decrease the initial temperature. The quark matter
contribution at large $p_T$ however drops as the initial time is increased
(the initial temperature increased).  

We note that the prompt photon production is about 17\% of the total 
yield. Noting that these are NLO results, the lowest order prompt
photon production is perhaps only of the order of 10\% of the total 
single photon production measured in the experiment. We also note 
that the thermal production of photons is almost identical to the 
prompt photon production when $\tau_0$ = 0.4 fm/$c$. 

We have also shown the results for the "Thermal+$\kappa$ $\times$ 
Prompt" photon contribution, with $\kappa$ adjusted to reproduce 
the experimental results at $p_T$ = 2.55 GeV/$c$. It is good to see 
that the same normalization provides a good description to the entire 
$p_T$ range in every case (see middle panels). We find that scaling 
the prompt photon results by factors of 2.7, 4.9, 5.4, 5.7, and 5.9, 
respectively are necessary in order to provide  a {\em quantitative} 
description of experimental results. 

We can perhaps argue that $\kappa$ accounts for the Cronin effect in 
case of nucleus-nucleus collisions as well as pre-equilibirum 
contributions which must surely be accounted for when $\tau_0$ 
is large. We do know that the pre-equilibrium electromagnetic 
radiations look thermal in nature~\cite{david}, and we have noted 
that in the present case, the prompt and the thermal contributions 
have similar slopes, for large initial temperatures

Even though a value of $\tau_0$ =0.2 fm/$c$ may be considered too small,
let us not forget that in this notation the nuclei would start interacting
at $\tau= -R/\gamma$ or  at about $\tau\approx -0.6$\,fm/$c$, and thus
a hot and dense system can be considered to be formed soon after the
complete overlap, which occurs at $\tau=0$ fm/$c$.

Let us try to see if some additional experimental result could 
actually distinguish between the different values for $\tau_0$,
and thus in a potentially interesting observation, we note (see 
lower panels of Figs.~\ref{fig5},\ref{fig6},\ref{fig7},\ref{fig8},
\ref{fig9}), that the elliptic flow parameter $v_2$ for the thermal 
photons~\cite{v2_phot} is quite sensitive to the formation time 
$\tau_0$~\cite{tau_0}. We also note the peak at  low $p_T$ in the 
$v_2(p_T)$, first noted by authors of Ref.~\cite{v2_phot} and 
interpreted as a consequence of competition of $\pi \pi \rightarrow 
\rho \gamma$ and $\pi \rho \rightarrow \pi \gamma$ reactions. We note 
that as we decrease $\tau_0$, the contribution of the quark matter 
increases. As this contribution arises from earlier times, where 
the momentum anisotropies are smaller, decreasing $\tau_0$ thus 
leads to an overall reduction in $v_2$ for thermal photons. Even 
though the azimuthal anisotropies for these fairly central collisions 
are small, they reveal an important sensitivity to the formation time
(see Fig.~\ref{fig10}. Note also the inversion of order of the results  
for $v_2$ with increasing $\tau_0$ with and without accounting for 
the non-thermal component). 

%Let us pause here to ponder over a question which has dogged the analysis
%of single photons from the WA98 experiment, from the very beginning.

Let us pause here to consider a question which has troubled
the analysis of single photons from the WA98 experiment, 
from the very beginning. In the present work, we have started 
with the assumption of a formation of QGP at time $\tau_0$. 
However, several studies have also~\cite{jane,pasi} presented 
a reasonable description of the data by assuming only the formation
of a hot hadronic gas in the collision without ever forming a QGP.
Which is the right scenario, then? We note here that the photon 
$v_2$ provides a possible resolution to this question. If no QGP 
is formed, then the $v_2$ for thermal photons will closely follow 
the $v_2(p_T)$ for $\rho$ mesons at larger $p_T$. Hence, it will 
be considerably larger than our prediction and also will rise 
monotonically~\cite{v2_phot} as $p_T$ increases. This suggests 
that a measurement of the $v_2$ of thermal photons along with 
their spectra could very firmly distinguish between the two
scenarios. 

Coming back to our present discussion,  we note that the results 
for $v_2$ for direct photons will be modified from the values for 
the thermal photons due to the presence of prompt photons (see 
Fig.~\ref{fig10}, lower panel). However, the {\em prompt photons 
as well as the pre-equilibrium photons} will not contribute to
the azimuthal anisotropy of the photon distribution, as they are 
not subjected to any collectivity. We can safely neglect the small
effect of azimuthal dependence of jet-quenching which may affect 
the fragmentation photons (which is less than about 30\% of the 
prompt contribution in the present case) or those of jet-induced 
photons~\cite{fms_phot}, which measure the anisotropy of the initial 
state~\cite{v2_gale}. This is because the QGP, if formed at the SPS, 
is very short lived and not very hot, as indicated by a small 
jet-quenching (not exceeding about 25--30\%) for such 
collisions~\cite{wa98_jet}. We show the results for final $v_2$ 
for single photons for the case of the WA98 experiment in Fig.~\ref{fig10}.

We finally recall a completely different calculation using the parton cascade
model~\cite{bms_phot1,bms_phot2} for this case, where the scattering and 
radiating partons produced a not-so-dense partonic system, but it was enough 
to reproduce the single photon production seen by the WA98 experiment beyond
about 3 GeV/$c$, {\it if} the partonic distributions are given an intrinsic
$< k_T >$ of about 0.44 GeV/$c$. In absence of the intrinsic $ < k_T >$ the
production is smaller by a factor of about 2. These photons can be considered
as due to prompt and the pre-equilibrium contributions.

We have so far assumed that the prompt and pre-equilibrium photons and 
their enhanced production due to intrinsic $k_T$ can be estimated by
 using a multiplicative factor $\kappa$ to the NLO pQCD results. Within 
this approach, we have ascertained that the photon observables,
especially the $v_2$, are sensitive measures of the initial condition.
Admittedly, there are some uncertainties in our approach such as the NLO
contribution to the thermal photon production and the effect of the 
viscous hydrodynamic evolution. At present, their effects are unknown 
although the effect of the finite viscosity on photons may soon be 
calculated\,\cite{Song:2008si}.

\subsection{Particle spectra}
How will the reported good
description of particle spectra~\cite{ksh} obtained using $\tau_0=$ 
0.8 fm/$c$, be affected, if a different value is used for $\tau_0$?
 Instead of discussing a complete calculation (with resonance decay
accounted for), we show the primary spectra of pions, rho mesons,
and protons for a typical impact parameter $b$ =7 fm, for different 
values of $\tau_0$, but keeping the entropy fixed as in the calculations 
discussed above (see Figs.~\ref{fig11},\ref{fig12},\ref{fig13}).
 We note that as the inverse slope for all the cases rises with
decrease in $\tau_0$ as the radial flow sets in earlier. We have 
checked that the increase in the inverse slope for pions is about 11\%, 
about 15\% rho mesons and protons as the initial time is decreased from 
1 fm/$c$ to 0.2 fm/$c$. We also note that the change in the spectra for 
the primary particles is quite marginal for $p_T$ below 1.5 GeV/$c$, 
even though it varies by a factor of about 3 at $p_T=$ 3 GeV/$c$. 
What is most interesting is that the differential elliptic flow 
parameter $v_2$ for hadrons is almost independent of the initial 
time. 

We conclude then, that the good description of  hadronic spectra
for low transverse momenta at SPS energies will remain unaffected by
the reduction of initial time from 1 fm/$c$ to 0.2 fm/$c$. This is
in contrast to what we saw earlier for thermal photons.

\section{Summary and conclusions}
We have re-analyzed the single photon production in $Pb+Pb$ collisions 
at the CERN SPS energies for 10\% most central collisions. Several 
improvements have been incorporated. The iso-spin, shadowing, and 
impact parameter dependence of the prompt photon production are 
explicitly included. NLO pQCD calculations are performed with the 
factorization, fragmentation, and renormalization scales fixed at 
$p_T/2$ based on a global description of the available data for 
$pp$ collisions. For the thermal photons calculations the initial 
conditions are taken as those which provided a good description to 
hadronic spectra, with a $\tau_0$ = 0.8 fm/$c$. We explored the 
consequences of using smaller initial times, keeping the entropy 
and the net-baryon number fixed.

We find that the data can be explained using a small formation 
$\tau_0$ of the order of 0.2 fm/$c$ when supplemented with prompt 
photons evaluated at NLO pQCD with a $\kappa$ factor $\approx$ 
2.7 to account for the Cronin effect. Larger initial times require 
much larger values for $\kappa$, which may be mimicking the 
pre-equilibrium contribution. A unique sensitivity to the 
formation time is seen in the photon elliptic flow, which 
could be useful in ascertaining whether a QGP was formed 
at the SPS energy.
 
\begin{acknowledgments}  
One of us (DKS) would like to acknowledge a very generous and warm 
hospitality at McGill University under the McGill India Strategic 
Research Initiative. We thank Charles Gale for useful discussions 
and valuable comments. The work of SJ~is supported in part by the 
Natural Sciences and Engineering Research Council of Canada.

The authors thank P.~Aurenche for providing the NLO pQCD code
and U.~Heinz for providing the hydrodynamics code.

\end{acknowledgments}

\end{document}